\documentclass[twoside,english]{elsarticlex}
\usepackage[T1]{fontenc}
\usepackage[latin9]{inputenc}
\usepackage{geometry}
\usepackage{float}
\usepackage{rotfloat}
\usepackage{textcomp}
\usepackage{amsbsy}
\usepackage{graphicx}
\usepackage{esint}
\makeatletter
\usepackage{color}
\usepackage{hyperref}
\usepackage{lineno}
\usepackage{ctable}
\usepackage{multirow}

\providecommand{\tabularnewline}{\\}

\makeatother

\usepackage{babel}
\begin{document}


\begin{frontmatter}{}


\title{Humidity Influence on Mechanics and Failure of Paper Materials: Joint Numerical and Experimental Study on Fiber and Fiber Network Scale}

\author[1]{Binbin Lin\corref{cor1}}

\ead{b.lin@mfm.tu-darmstadt.de}

\author[2]{Julia Auernhammer }

\author[3]{Jan-Lukas Sch\"afer }

\author[2]{Robert Stark }

\author[3]{Tobias Meckel}

\author[3]{Markus Biesalski }

\author[1]{Bai-Xiang Xu\corref{cor2}\fnref{}}

\ead{xu@mfm.tu-darmstadt.de}

\cortext[cor1]{Corresponding author}

\cortext[cor2]{Principal corresponding author}

\address[1]{Mechanics of Functional Materials Division, Institute of Materials
Science, Technische Universit\"at Darmstadt, Darmstadt 64287, Germany}

\address[2]{Physics of Surfaces, Institute of Materials Science, Technische Universit\"at
Darmstadt, Alarich-Weiss-Str. 16, 64287 Darmstadt, Germany }

\address[3]{Department of Chemistry, Macromolecular Chemistry and Paper Chemistry, Technische Universit\"at Darmstadt,
Alarich-Weiss-Str. 8, 64287 Darmstadt, Germany }

\begin{abstract}
Paper materials are natural composite materials and well-known to be hydrophilic unless chemical and
mechanical processing treatments are undertaken. The relative humidity impacts the fiber elasticity, the interfiber joint behaviour and the failure mechanism. In this work, we present a comprehensive experimental and computational study on the mechanical and failure behaviour of the fiber and the fiber network under humidity influence. The manually extracted cellulose fiber is exposed to different levels of humidity, and then mechanically characterized using atomic force microscopy, which delivers the humidity dependent longitudinal Young's modulus. We describe the relation and calibrate the data into a function of exponential form, and the obtained relationship allows calculation of fiber elastic modulus at any humidity level. Moreover, by using confoncal laser scanning microscopy, the coefficient of hygroscopic expansion of the fibers is determined. We further present a finite element model to simulate the deformation and the failure of the fiber network. The model includes the fiber anisotropy and the hygroscopic expansion using the experimentally determined constants, and further considers interfiber behaviour and debonding by using a humidity dependent cohesive zone interface model. Simulations on exemplary fiber network samples are performed to demonstrate the influence of different aspects including relative humidity and fiber-fiber bonding parameters on the mechanical features such as force-elongation curves, wet strength, extensiability.  Finally, fiber network failure in a locally wetted region is revealed by tracking of individually stained fibers using in-situ imaging techniques. Both the experimental data and the cohesive finite element simulations could demonstrate the pull-out of fibers and imply the significant role of the fiber-fiber debonding in the failure process of the wet paper.


\end{abstract}


\end{frontmatter}{}


\section{Introduction}

Cellulose-based fiber materials have been used since decades as packing, printing
media. Nowadays, they even have become popular as base material for electronic,
microfluidic devices on small scale \cite{gong2017turning,SHEN2019389,schabel2019role,liu2014filter}, and bio-composite \cite{pantaloni2021interfacial,regazzi2019microstructural,lee2006biodegradable} due to its recyclability to reduce pollution and save resources . 
As these devices are often exposed to a humid environment, their mechanical reliability and durability are often constraint and need to be understood before being used for massive application. The general understanding is that the mechanical property of natural composite-like paper-based materials is sensitive to the variation of relative humidity (RH). Salmen~et.~al.~\cite{salmen1980moisture} reported how RH affects the overall stiffness and tensile strength of the paper material. Upon increasing moisture
content at higher RH, paper material starts to exhibit a  ductile and more elastic behaviour, whereas upon drying the material becomes more brittle. However, the RH induced variation of the elastic property of single fibers and the influence of this variation on the overall paper sheet is still insufficiently characterized, as well as its influence on the load bearing and failure mechanism of the fiber joint and the fiber network. There are still controversial statements in the literature regarding the dependency of the fiber mechanical property on RH. Jajcinovic~et.~al.~\cite{jajcinovic2018influence}
has shown that the strength of individual fibers 
increases upon exposure to high RH, where as others have shown that the strength decreases. Based on single fiber testing at different RH, they reported that the breaking load of individual softwood fibers and fiber contacts displayed a maximum breaking load at 50\% RH, with the values showing a decreasing trend towards higher or lower RH. On the other hand, hardwoods show rather a decreasing trend in breaking loads at different RH levels. 
More importantly, the fiber-fiber bond strength is very sensitive to humidity. Dry paper strength is determined mostly by both the strength of their fibers and the interfiber bond strength between fibers. When dry paper breaks, the failure of both fibers and bonds has been observed. In the case of wet paper or with sufficiently high RH, the overall strength decreases. When it ruptures, less fibers break but mostly fiber bonds ~\cite{tejado2010does}. The interfiber joint strength mainly results from the hydrogen
bonds and depend on the fiber/fiber cross contacts, which can be broken by wetting of the fibers \cite{neimo1999papermaking}. The force holding the interfiber bonding is very sensitive to water, and the extent of bonding decreases with increasing water content~\cite{britt1948review,hubbe2008cellulosic}. Besides the loss of bond strength and decrease of elastic modulus, the swelling or hygroscopic expansion, which describes the moisture uptake, also plays an important role in RH sensitivity of the paper strength and extensiability. \cite{gamstedt2016moisture,joffre2013swelling,neagu2005influence,motamedian2019simulating} have reported that moisture adsorption due to humidity is the key feature to the dimensional stability loss of the cellulose-based paper structures. In order to improve the moisture resistance and identifying controlling parameters for engineering application, the underlying mechanisms were investigated by means of both experimental mechanics and numerical modelling techniques. Thereby moisture is typically characterised either in terms of RH in surrounding or the moisture content (relative moisture mass) in the specimen itself. The two measurements are intimately linked, and the relationship is characterised by the dynamic vapour sorption isotherms. While the RH is easily characterised by hygrometers in the ambient air, the determination of moisture content needs to weight the sample and compare it to its dry weight \cite{gamstedt2016moisture}. On the part of computational studies, a recent review by Simon \cite{simon2020review} extensively summarizes the numerical modelling approaches in study of material behaviour of paper and paperboard on different scales. Fiber network simulation are performed usually by means of direct simulation techniques, where individual fibers are explicitly modelled, and interfiber joint behavior considered by using a cohesive zone based modelling approach \cite{borodulina2018effect,Li:730091}. However, in the mentioned works, humidity dependent material parameters and models have not or not sufficiently been included, which remain an open topic for future research.

In the present paper, we combined numerical and experimental approaches to unveil systematically the RH impact on the mechanical behavior and the failure mechanism of paper materials on the fiber, and fiber network scale. Using atomic force microscopy (AFM) and confoncal laser scanning (CLSM) microscopy, we firstly characterized  the humidity dependency of the elastic modulus and the hygroscopic expansion coefficient (HEC) of  cellulose fibers. Instead of a raw fiber before the paper-making process, we studied  cellulose fibers manually extracted from a finished paper sheet. The determined parameters were then used as input in a finite element model to simulate the mechanical behavior and the failure of fiber networks. The force-elongation curves of the network gives an implication on the overall mechanical behavior. We studied particularly the mechanical features like the tensile strength, the effective stiffness and the extensiability. Additionally, by using a cohesive zone damage model, the fiber network simulations addressed the humidity dependent fiber-fiber contact bonding, which was resolved by the finite element cohesive interface elements. It allows the simulation of the local damage at individual fiber-fiber joints. At the same time, we evidenced single fiber pull-out during the tensile test of a locally wetted paper sheet, by labelling individual fibers and tracking their movement using in-situ imaging. We compared the cohesive zone finite element simulations and the experimental results, and confirmed the damage of the fiber network in a local wet state due to the interfiber joint separation.

\section{Experimental methods}

\subsection{Fiber sample and bending test}

For the bending experiments, the cellulose fibers were manually extracted from a cotton Linters paper sheet prepared according to DIN 54358 and ISO 5269/2 (Rapid-K\"othen process). The extracted fiber was mounted on a 3D printed sample holder, which supplied a fixed trench distance of 1 mm between the two attachment points. AFM was then performed by using a Dimension ICON (Bruker, Santa Barbara, USA) to measure static force-distance
curves along the cellulose fiber with a colloidal probe. The cantilever (RTESPA 525, Bruker, Santa Barbara, USA) with the nominal spring constant 200 N/m was modified with a 50 \textmu m colloidal probe (Glass-beads,Kisker Biotech GmbH \& Co. KG, Steinfurt, Germany) in diameter. All experiments were done in a climate chamber at the AFM. Hence, it was possible to vary the RH during the experiments. The chosen RH were 2 \%, 40 \%, 75 \% and 90 \%. As the RH was adjusted, the fiber was exposed 45 minutes to the environment before starting the measurements. For each fiber at each RH level, four different load levels with F = 500 nN, 1000 nN, 1500 nN, 2000 nN were applied and deflections of nine different segment points were evaluated.

\subsection{Determination of humidity dependant Young's modulus\label{sub:Determination-of-Young'smodulus}}

\begin{figure}[H]
\begin{centering}
\includegraphics[width=0.6\columnwidth]{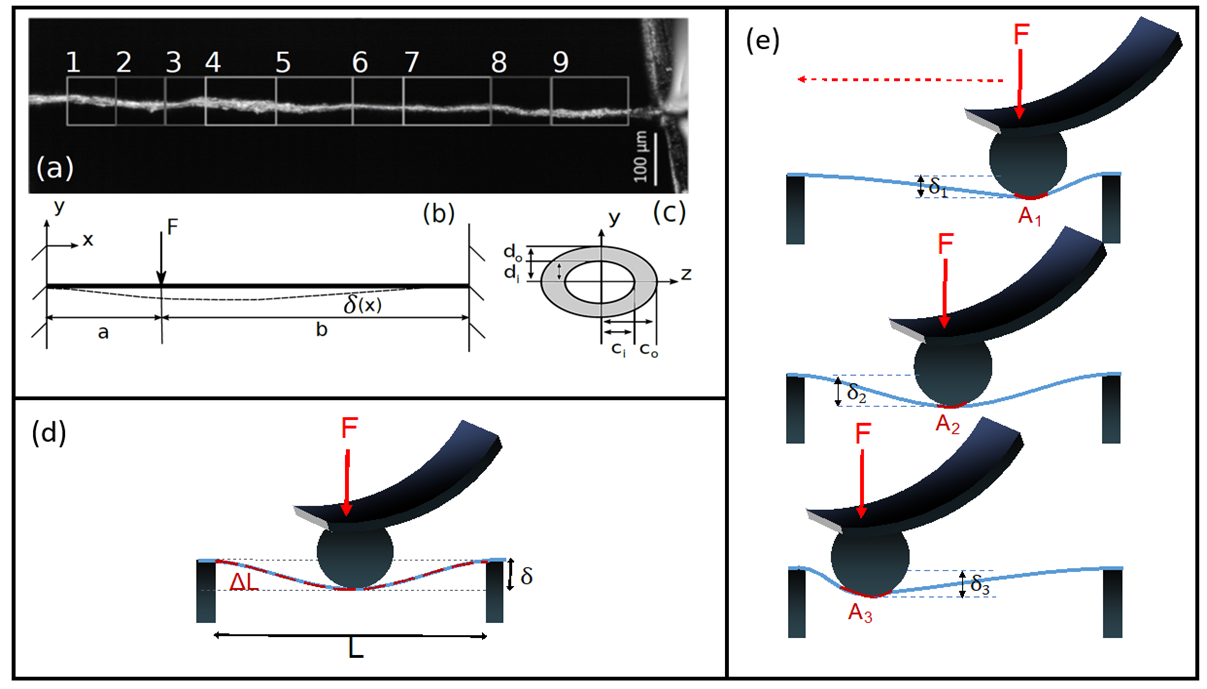}
\par\end{centering}

\caption{\label{fig:Mechanical Model setup} (a) An extracted fiber from a cotton Linters paper sheet, with the tapping segments framed and the cantilever end. (b) The corresponding mechanical beam
model. (c) The assumed hollow ellipse beam cross-section. (d) Standard three-point bending test method with the load applied to the middle of a tested sample. (e) Illustration of our performed scanning bending test, with aim to reduce local inhomogeneity afflicted results.}
\end{figure}

One extracted fiber sample is shown in Fig.~\ref{fig:Mechanical Model setup}(a). The fiber
was glued at both ends. Apart from standard bending experiments, we performed a scanning-like bending test, as we applied the load along the fiber length direction as illustrated in sub-figure (e). This aims to cover local inhomogeneity induced variations and give a proper statistical evaluation later on.

The fiber is modelled mechanically by using the beam model illustrated in Fig.~\ref{fig:Mechanical Model setup}(b). Based on the images, the fiber cross section is close to a hollow ellipse shown in Fig.~\ref{fig:Mechanical Model setup}(c). By use of the Euler beam theory, the differential equation governing the deflection $ \delta$ of the fiber and the corresponding solutions are given as:

\begin{eqnarray}
EI\cdot \delta''\left(x\right) & = & -M\label{eq:diff}\\
 \delta \left(x\right) & = & -\frac{Fb^{2}x^{2}\left(3a\left(a+b\right)-x\left(3a+b\right)\right)}{6EI\left(a+b\right)^{3}}\qquad\mbox{for \ensuremath{0\leq x\leq a}}\label{eq:Deflection}\\
 \delta \left(x\right) & = & \frac{Fa^{2}\left(a\left(a+b\right)-x\left(ab+a\right)\right)\cdot\left(a+b-x\right)^{2}}{6EI\left(a+b\right)^{3}}\qquad\mbox{for \ensuremath{a\leq x\leq a+b}}\label{eq:deflection}
\end{eqnarray}
where $\delta$ denotes the deflection, $E$ the longitudinal Young's modulus, $F$ the loading
force, $M$ the internal reaction moment, $I=\frac{\pi}{4}\left(c_{0}^{3}d_{0}-c_{i}^{3}d_{i}\right)$
the area moment of the assumed cross-section with $c_i, d_i$ and $c_o, d_o$ the main axes of the inner and of the outer ellipse, respectively. Eqs.~(\ref{eq:Deflection}) and (\ref{eq:deflection}) can be obtained
by calculating the internal reaction moment $M$ and integrating Eq.~\ref{eq:diff}. Then, using the boundary values $\delta(0)=\delta(a+b)=0$ and $\delta'(0)=\delta'(a+b)=0$ for the integration constants. Detailed procedure about solving statically indeterminate beam systems can be found in \cite{gross2009technische}. Using these equations, the average longitudinal Young's modulus $E$ is obtained by minimizing the difference between the measured and the calculated deflection from tapping different segments of the fiber via the least square approach: 
\begin{equation}
\sum\left(\delta\left(E\bar{I},\ x_{i}\right)-\delta_{i}\right)^{2}<tol
\end{equation}
where the positive constant $tol$ is the tolerance. One obtains, in fact, first the solution for $E\bar{I}$, and then determines the average $E$ along
the fiber sections by further
dividing the value $\bar{I}=mean(I)$, which denotes the averaged area moment
along the the fiber segments. This procedure was performed for every tested fiber at given load and RH level.

\subsection{Determination of hygroscopic expansion coefficient}\label{Determination of hygroscopic expansion coefficient}

A VK-8710 (Keyence, Osaka, Japan) confocal laser scanning microscope was used to investigate the swelling behaviour
of the fiber. In the climate chamber, where the RH was varied as in the AFM measurements previously, the fiber was suspended 45 minutes to the RH before starting the measurements. The swelling behaviour was analysed by the VK analyser software from Keyence (Osaka, Japan), after creating cross-section images along the fiber, where the local radii were measured. Further, in order to determine the HEC, the change in volume was calculated based on the experimentally recorded change of the cross-section at different RH. The reference volume of the non-prismatic fiber with assumed geometry was $V^{ref}=\pi\left(c_{o}d_{o}-c_{i}d_{i}\right)\cdot L$, see Fig. \ref{fig:Mechanical Model setup}(c). The total length of the fiber was assumed to be constant and do not vary with the changing RH, since both ends were clamped.  The relative change in volume due to RH can be then formulated as HEC multiplied by the absolute RH change $\Delta RH = RH - RH^{\mathbf{ref}}$ with $RH^{\mathbf{ref}}$ denoted as some reference RH state :

\begin{equation}
\Delta V/V^{ref}=\beta_{kk}\left(\Delta RH\right)
\end{equation}
where
\begin{equation}
\beta_{kk}=\beta_{kl}\delta_{kl}=\mathbf{trace} \left(\begin{array}{ccc}
\beta_{T} & 0 & 0\\
 & \beta_{T} & 0\\
\mbox{sym.} &  & \beta_{L}
\end{array}\right)=2\beta_{T}+\beta_{L}
\end{equation}
is the sum of 3D anisotropic hygroscopic expansion tensor $\beta_{kl}$ with in total 9 independent components in the general anisotropic case. Hereby $\delta_{kl}$ is the Kronecker-delta, $\beta_T$ is the HEC in the transverse direction, and $\beta_L$ HEC along the fiber length direction. In the transversely isotropic case, the hygroscopic expansion tensor has only the orthogonal components, namely those in the fiber longitudinal direction and in its cross-section. Further, it is experimentally validated considering the hierarchical layered wall structure~\cite{joffre2016method}, that the hygroscopic expansion in the fiber length direction, $\beta_{L}$, shown to be one order lower than that in the transverse directions, and therefore to be neglectable. Under this assumption, it leads to the following equation:

\begin{equation}\label{volumechange}
\frac{1}{2}\Delta V/V^{ref}=\beta_{T}\Delta RH
\end{equation}
Subsequently, the HEC in the transverse direction $\beta_{T}$ is calibrated with Eq.\ref{volumechange} at different $\Delta RH$ for every segment along the fiber length, as explained in Sec.~\ref{HEC section} later on.

\subsection{Paper sheet sample and tensile test \label{sub:Tensile-test}}

For studying the fracture mechanism of the paper materials on microscopic level, lab-engineered paper samples with bleached eucalyptus sulfate pulp (median fiber length (length-weighted): 0.80 mm; curl: 16.7 \%;
fibrillation degree: 5.2 \%; fines content: 8.0 \%) were used. The paper samples with grammages of 30 \textpm{} 0.6 g$\mathrm{m}^{-2}$ were prepared using a Rapid-K\"othen
sheet former according to DIN 54358 and ISO 5269/2. The tested samples have a gauge size of 30 mm in length and 15 mm in width. The standard procedure was slightly changed, in order to incorporate fluorescent labelled fibers. By
adding the dye (Pergasol Yellow F6-GZ liq. cationic dye) to the disintegrated fibers in the fiber suspension, the amount of labelled fibers in the paper samples was controlled. For the best results, 1 wt.-\% of the fibers were labelled with the dye. Afterwards the paper samples were conditioned for at least 24h under standard conditions (23 \textdegree C, 50 \% RH). 
A Zwick Z1.0 with a 20 N load cell using the software testXpert II
V3.71 (ZwickRoell GmbH \& Co. Kg) in a controlled environment with
23 \textdegree C and 50 \% RH was then used for tensile testing
with a constant strain-rate of 5 mm/min. In order to analyse the failure mechanism on a single fiber scale, a high magnification with a small
field of view was chosen. This was made possible by using a small
amount of distilled water to wet the paper samples in a defined region.
To analyze the the failure mechanism of the paper samples, a commercially
available full-frame mirrorless camera from Panasonic (DC S1) with
a macro lens from Canon (MP-E 65mm f/2.8 1-5x Macro Photo) and an
adapter from Novoflex (SL/EOS) was used. The camera was mounted on
a manual x/y/z-stage on a table that was decoupled from vibrations
of the tensile testing equipment. The aperture was set to 5.6, the
shutter speed was 1/30 sec and the ISO was set to 800. The videos
were recorded with a resolution of 3840 x 2160 px at a frame rate
of 29.97 frames/second. A UV-lamp (365 nm) was used from the backside
of the paper samples to excite the fluorophore of the labelled fibers.
In order to analyze the single images of the recorded videos, they
were converted from MP4 to AVI with an FFMPEG-script. Afterwards the
image-stacks in AVI format are processed using the program Fiji \cite{schindelin2012fiji}. Processing
included cropping of the images, so that only the part of the failing
fiber network  is visible, extracting the green-channel (where the fluorescing
fibers are most visible) and saving the image-stack in an uncompressed
TIF-format for further analysis.

\section{Computational simulations using the cohesive zone finite element model}

\subsection{Mechanical model of the single fiber}

We apply the linear elasticity to the fibers, including the stress equilibrium, the linear kinematics and the transversely istropic linear elastic material law:   
\begin{eqnarray}
\sigma_{ij,j} & = & 0\\
\varepsilon_{kl} & = & \frac{1}{2}\left(u_{k,l}+u_{l,k}\right)\\
\sigma_{ij} & = & C_{ijkl}\left(\varepsilon_{kl}-\varepsilon_{kl}^{h}\right) = C_{ijkl}\left(\varepsilon_{kl}-\beta_{kl}\Delta RH\right)
\end{eqnarray}
in which $\sigma_{ij}$ is the Cauchy stress tensor, $C_{ijkl}$ the stiffness tensor and $\varepsilon_{kl}$ the strain tensor. The strain due to the hygroscopic
expansion is given as $\varepsilon_{kl}^{h}=\beta_{kl}\Delta RH$,
where $\beta_{kl}$ and $\Delta RH$ denote the anisotropic HEC tensor as described in the previous section and the relative humidity change, respectively.  The transversely isotropic constitutive material law is applied for the fiber anisotropy along the fiber direction. Thus the inverse of the stiffness tensor can be given as follows:
\begin{eqnarray}
C_{ijkl}^{-1} & = & \left[\begin{array}{cccccc}
1/E^{T} & -\nu^{TT}/E^{T} & -\nu^{LT}/E^{L}\\
-\nu^{TT}/E^{T} & 1/E^{T} & -\nu^{LT}/E^{L} &  & \mbox{Sym}.\\
-\nu^{LT}/E^{T} & -\nu^{LT}/E^{T} & 1/E^{L}\\
 &  &  & 1/G^{T}\\
 & 0 &  &  & 1/G^{T}\\
 &  &  &  &  & 1/G^{L}
\end{array}\right]
\end{eqnarray}
with five independent parameters: $E^{L}$,\textbf{ $E^{T}$}, $G^{T}$
the longitudinal modulus, the transverse modulus, the shear modulus and $\nu^{LT}$, $\nu^{TT}$ the two Poisson's ratios. The values of the Poisson's ratios are specified in Tab.~\ref{tab:elastic parameters}. 
Base on \cite{magnusson2013numerical}, one can further reduce the number of elastic parameters, by assuming the correlations between other elastic parameters and the longitudinal Young's modulus $E^{L}$ with the unified property-related S2-Layer. This assumption is based on the fact that this layer represents the main constituent of the fiber. The relations between the elastic parameters shown in Tab.~\ref{tab:elastic parameters} are employed in the following finite element simulations. Note that the longitudinal Young's modulus $E^{L}$  and the HEC are experimentally determined as explained in Sec.~\ref{sub:Determination-of-Young'smodulus}. Further, on the single fiber level, no fiber damage is assumed.

{\renewcommand{\arraystretch}{1.5}
\begin{table}[H]
\caption{The elastic parameters and their dependency on the longitudinal Young's modulus $E^{L}$.~\cite{magnusson2013numerical} }
\label{tab:elastic parameters}

\centering{}%
\begin{tabular}{ccccccc}

\hline 
Elastic parameters & $E^{L}$ & $E^{T}$ & $G^{T}$ & $G^{L}$ & $\nu^{LT}$ & $\nu^{TT}$ \tabularnewline 
\hline
Value  & $E^{L}$ & $E^{L}$/11 & $E^{L}$/23 & $E^{T}/2(1+\nu^{TT})$ & 0.022 & 0.39\tabularnewline
\hline 
\end{tabular}
\end{table}}

\subsection{Cohesive zone interface model for fiber-fiber contact in the dry state}

Similar to works \cite{Li:730091,magnusson2013numerical}, a cohesive
zone-based approach is utilized to characterize the debonding behavior. See Fig. \ref{fig: FEM Condition and CZM}(d) for the illustration. 
In the current work, we applied a non-potential based CZM \cite{mcgarry2014potential},
which aims to give a proper behavior in mixed-mode loading scenario
to avoid the fiber penetration. The traction-separation law are given as:
\begin{eqnarray}
\label{Eq:TS-law}
T_{n}\left(\Delta_{n},\Delta_{t}\right) & = & \sigma_{max}\exp\left(1\right)\left(\frac{\Delta_{n}}{\delta_{n}^c}\right)\exp\left(-\frac{\Delta_{n}}{\delta_{n}^c}\right)\exp\left(-\frac{\Delta_{t}^{2}}{{\delta_{t}^c}^2} \right)\\
T_{t}\left(\Delta_{n},\Delta_{t}\right) & = & \tau_{max}\sqrt{2\exp\left(1\right)}\left(\frac{\Delta_{t}}{\delta_{t}^c}\right)\exp\left(-\frac{\Delta_{n}}{\delta_{n}^c}\right)\exp\left(-\frac{\Delta_{t}^{2}}{{\delta_{t}^{c}}^2}\right)
\end{eqnarray}
where $T_{n},\ T_{t}$ are the traction components of $\mathbf{T}$
in their normal and tangential loading state. $\Delta_{n},\ \Delta_{t}$
are the displacement separation at the fiber-fiber bonds. $\sigma_{max}$
and $\tau_{max}$ are the maximal stresses in pure normal and pure
shear separation, $\delta_{n}^{c}$ and $\delta_{t}^{c}$ the critical
length to the maximum stresses, respectively. The energy per surface
for both separation modes are given with $\phi_{n}=\delta_{n}^{c}\sigma_{max}\exp\left(1\right)$
and $\phi_{t}=\delta_{t}^{c}\tau_{max}\sqrt{0.5\exp\left(1\right)}$.
The damage variable $D$ is defined as:

\begin{equation}
D=\frac{\Delta_{eff}-\Delta_{c}}{\Delta^{f}-\Delta_{c}}
\end{equation}
ranging from 0 to 1 and represents the intact state and fully damaged state.  $\Delta_{eff}=\sqrt{\Delta_{n}^{2}+\Delta_{t}^{2}}$, $\Delta_{c}=\sqrt{\left(\delta_{n}^{c}\right){}^{2}+\left(\delta_{t}^{c}\right){}^{2}}$
are the effective mixed-mode separation and the separation corresponding
to the maximum stresses, and $\Delta^{f}=\sqrt{\left(\delta_{n}^{f}\right)^{2}+\left(\delta_{t}^{f}\right)^{2}}$
 the final separation at complete failure. ~Eq. \ref{Eq:TS-law} and 13 are referred as the traction-separation law in the dry state $\mathbf{T}^{dry}$ for the following humidity dependent cohesive zone model.

\subsection{Humidity dependency of the cohesive zone model \label{Humidity dependency of the cohesive zone model}}

\begin{figure}[H]
\begin{centering}
\includegraphics[width=1.0\columnwidth]{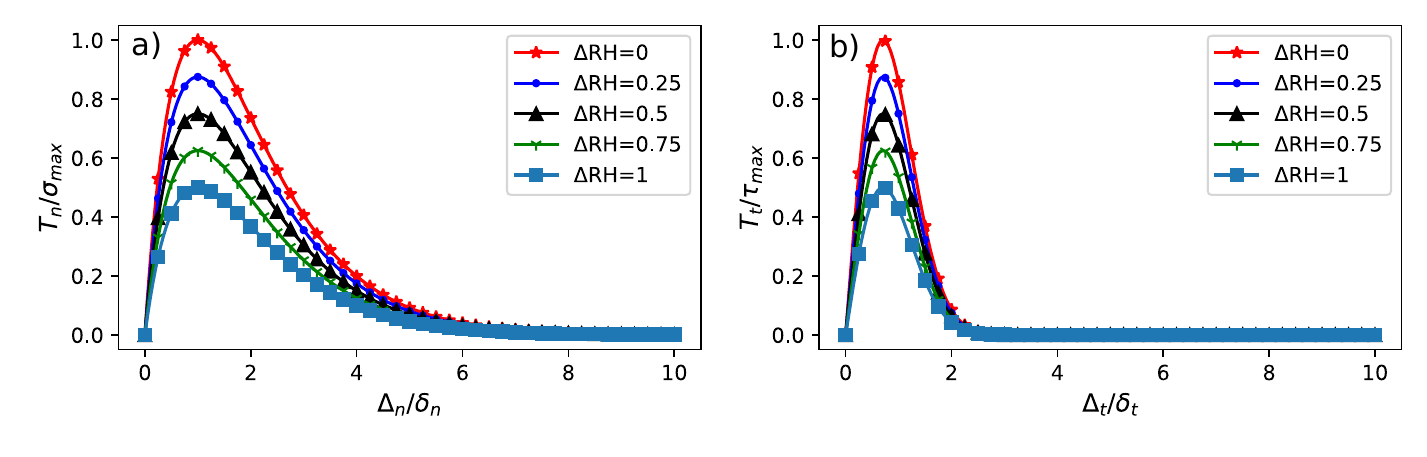}
\par\end{centering}

\caption{\label{fig:Cohesive law variation}Decay of traction-separation law
due to RH with K = 0.5 (a) Normal cohesive behaviour (b) Tangential
cohesive behaviour }
\end{figure}

In a subsequent step, the cohesive zone model due to the humidity
influence is modified by multiplying a decrease term:
\begin{eqnarray}
\label{CZhumidity}
\mathbf{T}\left(\Delta RH\right) & = & \mathbf{T}^{dry}\left(1-K\left(\Delta RH\right)\right)
\end{eqnarray}
This is motivated by \cite{jemblie2017review}, mainly based on
using CZM for hydrogen embrittlement of steel structures, where the
effect of hydrogen on the accelerated material damage are studied.  Here, the analogy and simple assumption is understood as the cohesive strength decreases with increasing RH, where the separation distance remains the same. $K$ is a softening parameter and describes the degree of decrease and the bound of the allowable cohesive strength, which needs to be adjusted, when experimental data are available. This intuitive assumption is made for the first step, since there are barely experimental data in the literature regarding RH dependency of the mechanical property of interfiber joints. Jajcinovic et al. \cite{jajcinovic2018influence} tested the strength of interfiber joints for softwood and hardwood interfiber joints at three RH levels. While for cases of hardwood, the breaking strength decreases, for cases of softwood fibers, the optimal breaking strength seems to favor at 50\% RH, showing decreasing trend towards higher or lower RH values. However, a statistically sound conclusion cannot be drawn, since a larger variation of breaking load are observed for the softwood joints. Furthermore, hardly any literature on RH-dependent cohesive separation parameters could be found. Therefore, as a first step towards considering RH dependent cohesive relation, we make use of this assumption. Without loss of generality, this relation can be improved upon available, accurate experimental results.
 For a $K=0.5$, the traction-separation function are plotted in Fig. \ref{fig:Cohesive law variation}.

\subsection{Weak formulation for cohesive FE implementation}

The FE implementation is based on the governing equations describing the general bulk deformation and the cohesive zone damage model as given previously. The corresponding weak formulation can be obtained from the principle of virtual
work based on~\cite{park2012computational} but with extension to include further the cohesive interface contributions:

\begin{equation}
\int_{\Omega}\delta\mathbf{\boldsymbol{\varepsilon}}:\mathbf{\boldsymbol{\sigma}}\ \mbox{d}V+\int_{\Gamma_{int}}\delta\mathbf{\Delta}\cdot\mathbf{T}\ \mbox{d}S=\int_{\Gamma}\delta\mathbf{u}\cdot\mathbf{T}_{ext}\mbox{d}S\label{eq:Weak form governing equation-1}
\end{equation}
where $\delta\mathbf{\boldsymbol{\varepsilon}},\delta\mathbf{u},\delta\mathbf{\boldsymbol{\Delta}}$
are the virtual strain tensor, the virtual displacement vector and the virtual separation vector at the interface,
respectively. $\mathbf{\boldsymbol{\sigma}}$ is the tensor notation of the Cauchy stress tensor $\sigma_{ij}$,
$\mathbf{T}$ is the traction vector at interface $\Gamma_{int}$ with the normal and tangential components $T_n, T_i$
and $\mathbf{T}_{ext}$ the external traction on outer boundary $\Gamma$.
Eq.~\ref{eq:Weak form governing equation-1} describes the energy
balance between the strain energy or the work of the internal force in the domain
$\Omega$ and the work contributed by the interface elements and
by external forces.
The non-potential based cohesive zone model is implemented in an user material kernel in the open source FE software MOOSE~\cite{permann2020moose}. A detailed step-by-step instruction of general cohesive FE implementation from Eq. \ref{eq:Weak form governing equation-1} can be found in \cite{park2012computational}.

\subsection{Fiber network generation and FE simulation setups\label{fiber network generation and FE simulation conditions}}

\begin{figure}[H]
\begin{centering}
\includegraphics[width=1.0\columnwidth]{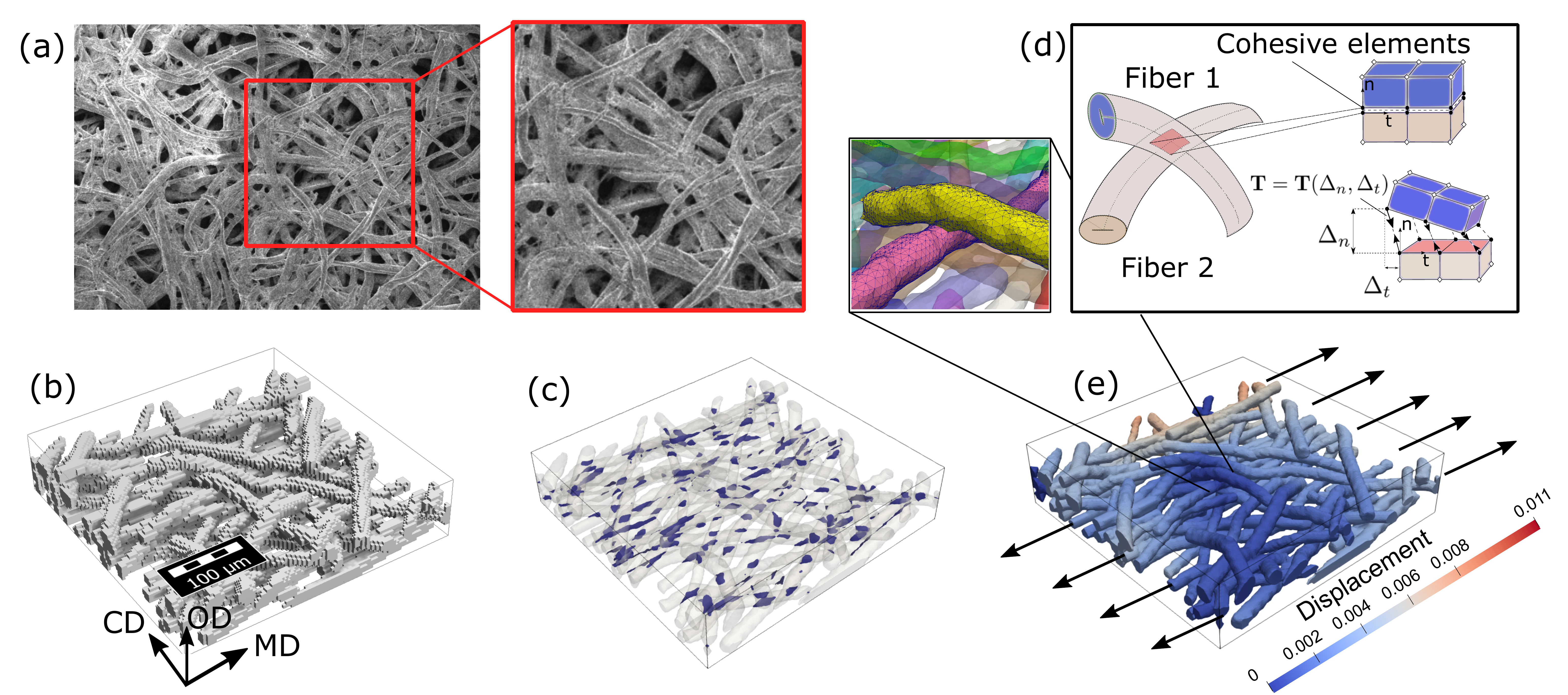}
\par\end{centering}
\caption{\label{fig: FEM Condition and CZM} (a) SEM image showing the fiber network of the cotton Linters paper sample. (b) Synthetic fiber network generated by Geodict in voxelized and unrendered form. (c) Visualization of the contact areas inside the fiber network (d) Illustration of the cohesive zone model for the interfiber contact (Hexahedron elements are used for a simple visualization) (e) The setup of the cohesive finite element simulation with the prescribed displacement along the machine direction (MD) and figure insert showing the FE-mesh for two fibers that are in contact.}
\end{figure}

The Software Geodict$\textsuperscript{\textregistered}$ was used
to create periodic, synthetic fiber network samples, which permits a similar deposition procedure as reported in \cite{kulachenko2012direct,Li:730091}. It uses voxels to represent the structure of the fiber network after the structure is generated for the given geometric parameters. Upon initialization, a voxel size of one micron was chosen, see Fig.~\ref{fig: FEM Condition and CZM}(b). The fibers are deposited by assuming Gaussian distributions of fiber orientation, diameter and length. For simplicity, we neglected the complex shape of the cross-section and assumed that the fiber is solid and has a circular-shaped cross section along the
fiber length. They were created in a flat-plane, afterwards fell under gravitational force, and were then elastically deformed and deposited onto the previous fibers. The deposition process was completed when the specified grammage was reached. 
After generation of the voxel geometry, the surface mesh was created. Afterwards, coarsening and smoothing step were performed until sufficiently fine surface mesh was obtained. The surface mesh was then exported and meshed to volume elements using the open source meshing software Gmsh \cite{geuzaine2009gmsh}. The modeled fiber network sample included approximately 1 million linear tetrahedron
volume elements. About 40000 local 2D-elements were used to resolve the contact area in order to achieve mesh independence. Details on the choice of the mesh size is given in the supplementary information. For the present work, the fiber network setting and features are given in 
Tab.~\ref{tab:fiber network setting}. The size of the simulated fiber network
sample is similar to that used in~\cite{Li:730091} and our previous work \cite{lin2020data}. The diameter of the fiber is similar to our tested fiber. Due to the periodic fiber network structure, the length of the fibers in the simulation is chosen to be around half of the median fiber length (length-weighted), that were tested from the paper sheet as mentioned previously.   For the finite element simulation, we calculated the orientation tensor for each fiber in the voxel-based
geometry by use of Euler angles. The orientation tensor was then used for transformation of the elasticity tensor for each fiber. Regarding the boundary conditions, $u_{MD}=0$ is specified on the left boundary, $u_{MD}=u$ on the right boundary and $u_{CD}=0$
on the front and back boundary.  These correspond 
to displacement boundary condition in the machine direction (MD) with the boundary cross-section kept as a plane.   The out-of-plane (OD) degree of freedom and others on each boundary side were unspecified and stress-free. CD is hereby referred as the cross-machine direction. FEM simulations
were performed on a high performance computer with 24 cores for around 8 hours.

{\renewcommand{\arraystretch}{1.1}
\begin{table}[H]
\caption{\label{tab:fiber network setting} Geometry parameters of the generated periodic fiber network sample. }
\centering{}%
\begin{tabular}{cc}
\hline 
Parameters & Values\tabularnewline
\hline 
Domain size x, y  & 400 $\mu\mathbf{\mbox{m}}$\tabularnewline
Distribution  & Gaussian\tabularnewline
fiber network orientation & 0\textdegree{} of MD \tabularnewline
Fiber length & 300 $\mu\mathbf{\mbox{m}}$\tabularnewline
Fiber diameter  & 17 $\mu\mathbf{\mbox{m}}$\tabularnewline
Grammage & 30 $\mbox{g/\ensuremath{\mbox{m}^{2}}}$\tabularnewline
Standard deviation (Std) orientation & $\pm$36\textdegree{} of MD \tabularnewline
Std fiber length & $\pm$18 $\mu\mathbf{\mbox{m}}$\tabularnewline
Std diameter  & $\pm$2.2   $\mu\mathbf{\mbox{m}}$\tabularnewline
\hline 
\end{tabular}
\end{table}}

\section{Results and discussion}

\subsection{Parameterization of humidity dependent elastic modulus}

\begin{figure}[H]
\begin{centering}
\includegraphics[width=0.9\columnwidth]{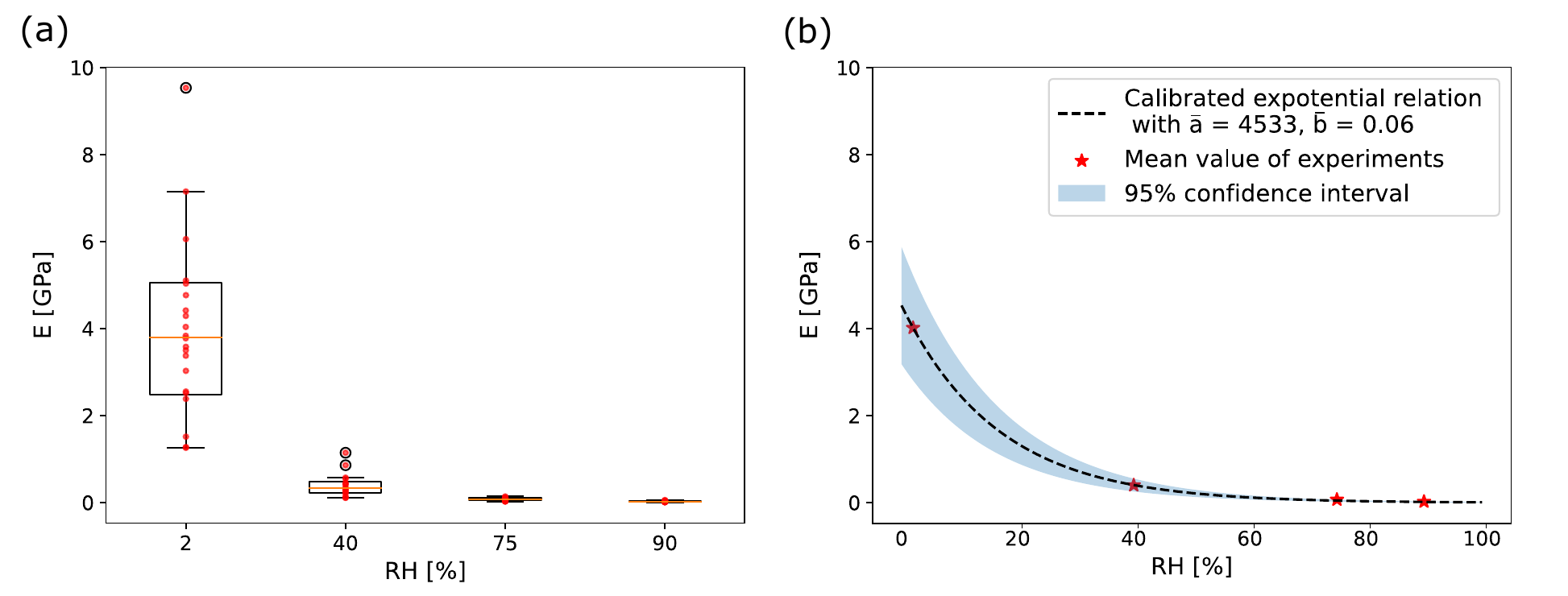}
\par\end{centering}

\caption{\label{fig:E-RH value} (a) Distribution of Young's modulus data in dependency of RH. In total, 20 data points were evaluated for 5 fibers and 4 applied loads, at given RH levels.  (b) Calibrated exponential law with corresponding mean values as star dots and its 95\% confidence interval.}
\end{figure}

The experimentally determined values using aforementioned technique of the longitudinal Young's modulus $E^{L}$ for the prescribed loading
at different RH values are presented in form of boxplot in Fig.~\ref{fig:E-RH value}(a). For simplicity, $E^{L}$ will be written as E in the following. For each box, Q1 to Q3 quartile values of the data, (corresponding to 25\% and 75\%) are drawn inside the box. The yellow line inside the box marks the median (Q2). The whiskers extend from the edges of box to show the range of the data. The extension of the data are limited to 1.5 * IQR with IQR = Q3 - Q1 from the edges of the box, ending at the farthest data point within that interval. Outliers are plotted as separate black framed dots.
From the boxplot, it is clear that E decreases with increasing RH levels. This is in agreement with the literature.
The Young\textquoteright s modulus reduces because water molecules act as lubricant between the cellulose molecules, as the water molecules
infiltrate the network and destroy the hydrogen bonds between the cellulose molecules \cite{blechschmidt2013taschenbuch, janko2015cross,quesada2011nanomechanical}.
Further, it is noticeable that for  RH$ = 2\%$, a higher variation is observed due to distribution of determined data points. Apart from one outlier, the median is obtained to be approximately 4 GPa. The variation decreases with increasing RH level, which is given by the shrinkage of the boxes. Compared to the values available in the literature, the obtained elastic modulus is few times lower. Fernando et al. \cite{fernando2017mechanical} obtained from an AFM-based three point bending
test a Young's modulus of 24.4 GPa of softwood fibers. From tensile tests of single fibers,
a range from 20-80 GPa has also been reported in~\cite{eichhorn2001young, kompella2002micromechanical}. There is one exception~\cite{miyake2000tensile}, which reported a Young's modulus in the MPa range. 
The lower level of our Young' modulus can be attributed to the fibers under testing. As stated previously, we manually extracted the single fibers from a finished paper sheet. Thus, our fiber was a part of the completed paper after production process with beating or pressing. In other words, the fiber tested is not anymore in its natural raw state. In fact, in the paper making process, the layered wall structure of the fiber is often milled off, which could lead to softening of
the fiber. Furthermore, a small degree of determined 
Young's moduli seem to increase with the applied
load. This may stems from nonlinear elasticity or nonlinear geometrical deformation which are neglected in the current mechanical models. Further, we propose the Young's modulus E as a function of RH and approximate the decaying due to the humidity effect in an exponential form. Afterwards, the coefficients of the exponential function are calibrated to the mean values that we determined by the least square approach from the experimental bending tests. The exponential function takes the following form:
\begin{equation}
E(RH)=\bar{a} \cdot e^{-\bar{b}RH}
\label{eq:Polynomial}
\end{equation} where $\bar{a}$ can be understood as $E_{max}$, which represents the averaged modulus extrapolated at 0\% RH and to be around 4.5 GPa. $\bar{b}$ determines the degree of elastic decaying. The calibrated coefficients $\bar{a}$, $\bar{b}$
are given in Tab.~\ref{tab:E-RH-coefficient}. The E-RH curve is depicted in Fig.~\ref{fig:E-RH value}(b), along with the deviation as 95 \% confidence interval. With the formula given in Eq.~\ref{eq:Polynomial}, Young's modulus can be determined at any RH values and can be used further for simulation of RH-dependent material behaviour.

\begin{center}
{\renewcommand{\arraystretch}{1.3}
\begin{table}
\caption{\label{tab:E-RH-coefficient} The determined coefficients for the RH dependency of the longitudinal Young's modulus defined in Eq.(\ref{eq:Polynomial}).}

\centering{}%
\begin{tabular}{ccccc}
\hline 
Coefficient & $\bar{a}$ & $\bar{b}$ &  \tabularnewline
\hline 
Value & 4533 & 0.06 & \tabularnewline
\hline 
\end{tabular}
\end{table}}
\par\end{center}

\newpage
\subsection{Hygroscpic expansion coefficient}\label{HEC section}

\begin{figure}[H]
\begin{centering}
\includegraphics[width=1\columnwidth]{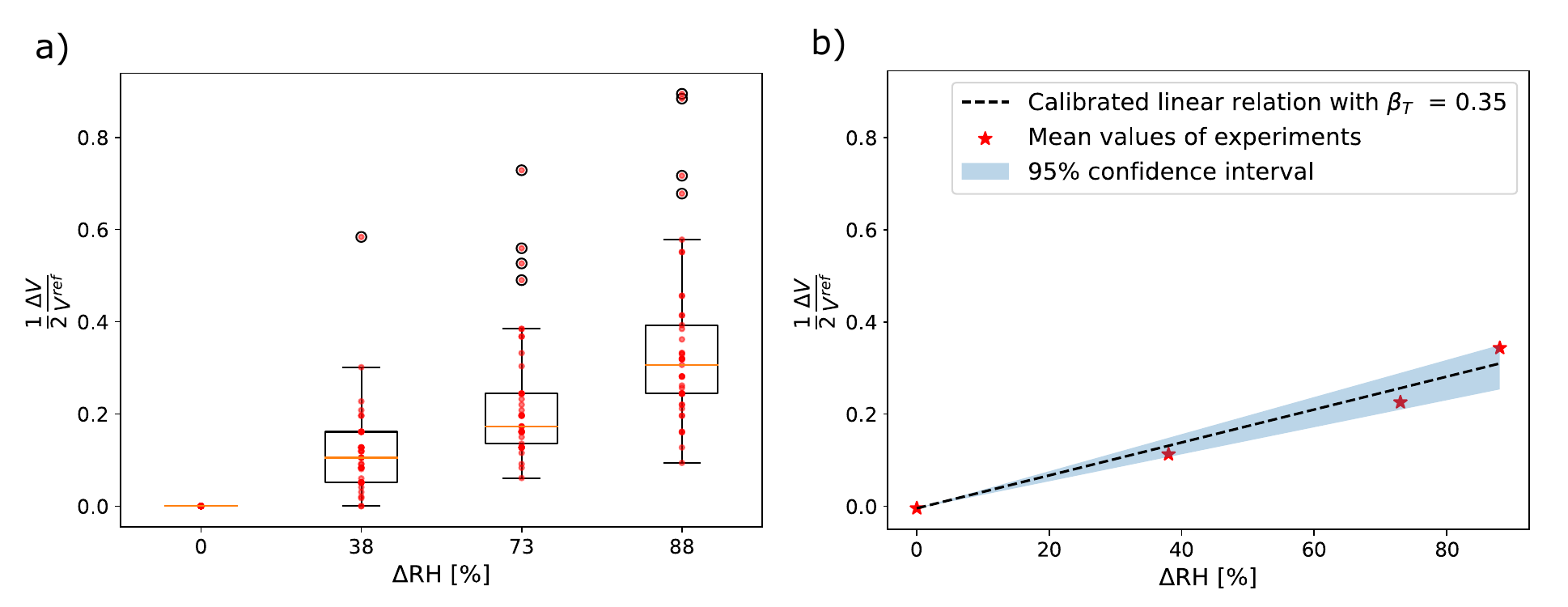}
\par\end{centering}

\caption{\label{fig:HygroCoefficient} (a) Distribution of measured volumetric strain data. In total 45 data points were evaluated, with data from 9 segments per fiber at each RH change. (b) Calibrated linear relation of volumetric strain with $\Delta$RH and the corresponding mean values as star dots, and its 95\% confidence band. The slope of the linear curve is then determined as the transverse HEC.}
\end{figure}

The experimentally determined volume change of different segments of a fiber exposed to different level humidity are summarized in Fig.~\ref{fig:HygroCoefficient}(a). Together, data of five fibers, each of nine segments were collected and 45 data points were evaluated in total.  Due to the high local inhomogeneity, high deviation of the data points for different segments are observed, which leads to the larger distribution in the boxplot at each RH change. $\Delta$RH is referred to the difference between the current RH level and reference RH state, which is 2\% RH. Similar to the determination of E-RH relation, the HEC is determined in Fig.~\ref{fig:HygroCoefficient}(b). The volumetric strains are averaged over the data points at each $\Delta$RH level.  The averaged volume strain assumes a linear dependency on the relative humidity, based on the relation Eq.~(\ref{volumechange}). The transverse HEC $\beta_{T}$ can be thus extracted as the slope of the curve, which is indicated by the dashed black line, and the value is determined as 0.35. This is in good agreement with the values reported in \cite{joffre2013swelling} as collected by \cite{joffre2016method} from different data available in the literature, ranging from 0.2 - 0.44. It should be noted that the recorded outliers in the boxplot originate from the segments that contained inflated microfibrils, as e.g segment number 4 in Fig.~\ref{fig:Mechanical Model setup}(a),  which would bias the overall results. There are several other methods in determining the HEC reported in \cite{joffre2016method}, such as measuring from X-$\mu$CT or back-calculation from composite properties as in \cite{neagu2007modelling,almgren2009moisture}. Since our focus is on the influence of RH on mechanical behaviour of the fiber network, the readers are referred to literature for further investigations about the dimensional stability loss in relation with HEC, see e.g. \cite{neagu2005influence}.

\newpage{}
\subsection{Simulation of force-elongation relation on fiber network scale}
\begin{figure}[H]
\begin{centering}
\includegraphics[width=0.75
\columnwidth]{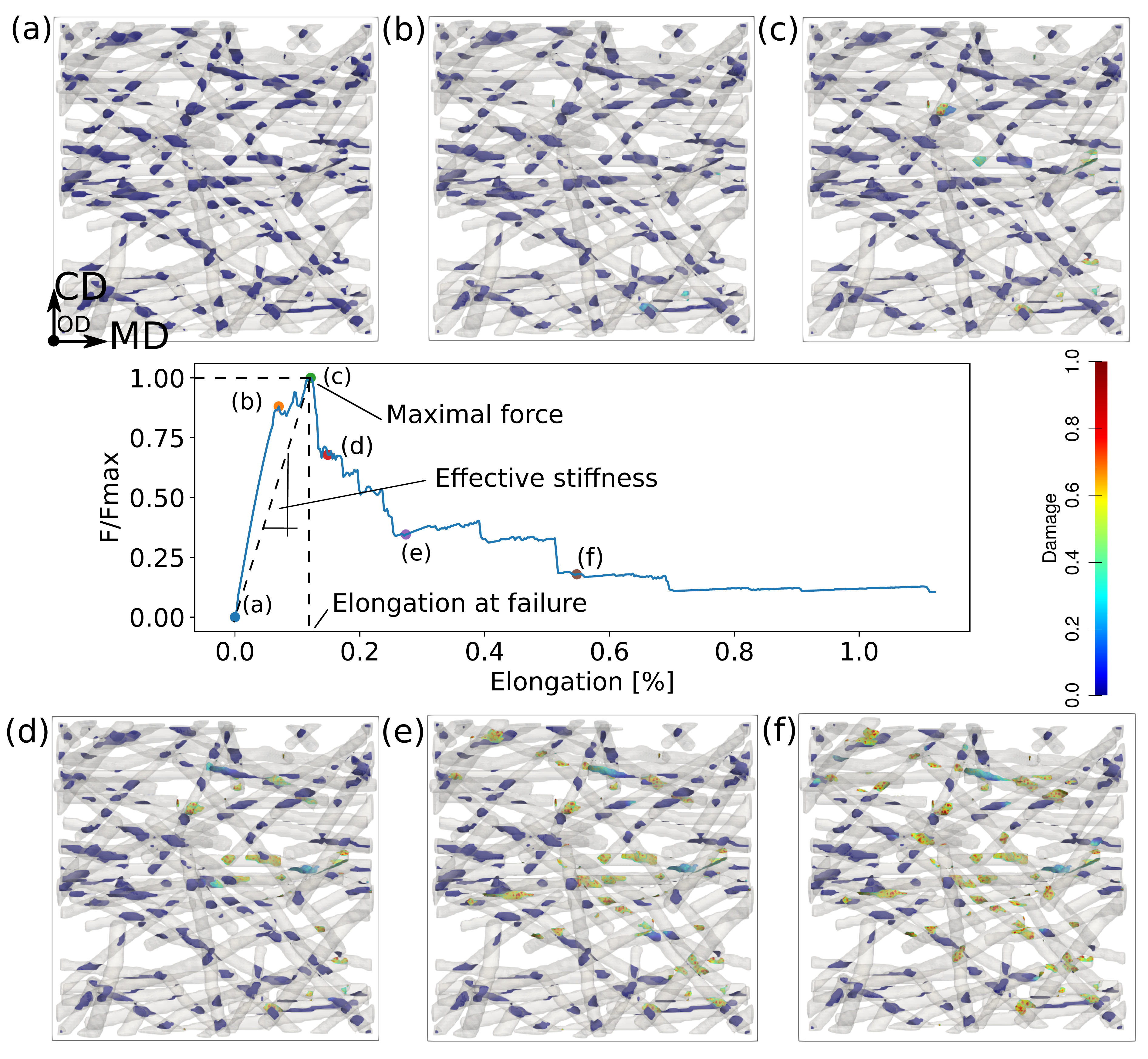}
\par\end{centering}

\caption{\label{fig:Coloured Fibers and Damaged} Deformation and damage process
of an exemplary fiber network under the tensile loading along the MD direction. (a-f): The snapshots of deformed fiber network for the corresponding points in the force-elongation curve shown in the middle. The damage of the fiber-fiber contact is indicated by the color count plot.}
\end{figure}

Bulk and cohesive finite element simulations were carried out on the generated fiber network under the prescribed boundary conditions in Sec.~\ref{fiber network generation and FE simulation conditions}. One fiber network sample without humidity influence is shown for demonstration purpose in Fig.~\ref{fig:Coloured Fibers and Damaged}(a), with the cohesive fiber-fiber interface highlighted in dark blue. The sample deforms under the tensile displacement along the MD. A few snapshots on the deformation and the interface damage level are sequentially presented in Fig.~\ref{fig:Coloured Fibers and Damaged}(b-f). We simultaneously documented the normalized reaction force (w.r.t maximal value of the reaction force) and the elongation of the sample. The corresponding curve is depicted as the insert of the figure.

 The points which correspond to the snapshots (a-f) are indicated on the curve. It can be seen that mechanical response from
(a) to (b) is almost linear. From (b) to (c), damage initiation occur at certain areas. The maximal force reaches at (c), and from (c) to (d), the force drops about 35\%. The force at point
(c) is referred as the maximal force or the failure force, while the corresponding elongation is termed as elongation at failure in the following context. From (c) on, damage processes continue more extensively. In particular, from (e) and (f) almost all fiber contacts are damaged, and the sample bears hardly any load (20\% of the maximal force) due to the damage percolation. The numerical convergence at this stage becomes difficult for implicit finite element solver due to the large deformation and the energy release stored in the fiber bonds. 

Using the established fiber network simulation framework, we firstly carried out a series of material parameter studies without humidity influence. It is necessary to understand the extent of the influence of those potentially important material parameters on the mechanical behavior, because the material parameters inherently varies with  different kind of fibers. Simulations with humidity influence are reported thereafter. It should be noticed that for the following parameter studies, force-elongation curve up to force at failure and the corresponding elongation at failure is considered, as indicated by point (c) in Fig.~\ref{fig:Coloured Fibers and Damaged}. Force-elongation behaviour in post-damage range are out of our current scope, where the numerical convergence may be difficult for implicit finite element solver as previously mentioned. Tab.~\ref{tab:Parameter-study} summarizes the default values and the corresponding variation sets of the material parameters considered. Unless it is otherwise stated, only one set of the material parameters is varied at once, while the other parameter sets are assumed to take their default values.

{\renewcommand{\arraystretch}{1.2}
\begin{table}[H]
\caption{\label{tab:Parameter-study}Value for parameter study without humidity influence}

\centering{}%
\begin{tabular}{cccc}
\hline 
Parameters & Symbol & Default values & \multicolumn{1}{c}{Variations}\tabularnewline
\hline 
Cohesive strength & $\sigma_{max},~\tau_{max}$ {[}MPa{]}  & 1,~0.25 & 0.1,~0.025;~ 10,~2.5;~ 20,~5 \tabularnewline
Final separation & $~\delta_{t}^{f}, \delta_{n}^{f}$ {[}$\mu\mbox{m}${]} & 1 & 0.1; 5; 10\tabularnewline
Young's modulus & E {[}GPa{]} & 4.5 & 0.45; 45; 90 \tabularnewline
\hline 
\end{tabular}
\end{table}}

\begin{figure}[]
\begin{centering}
\includegraphics[width=0.5\columnwidth]{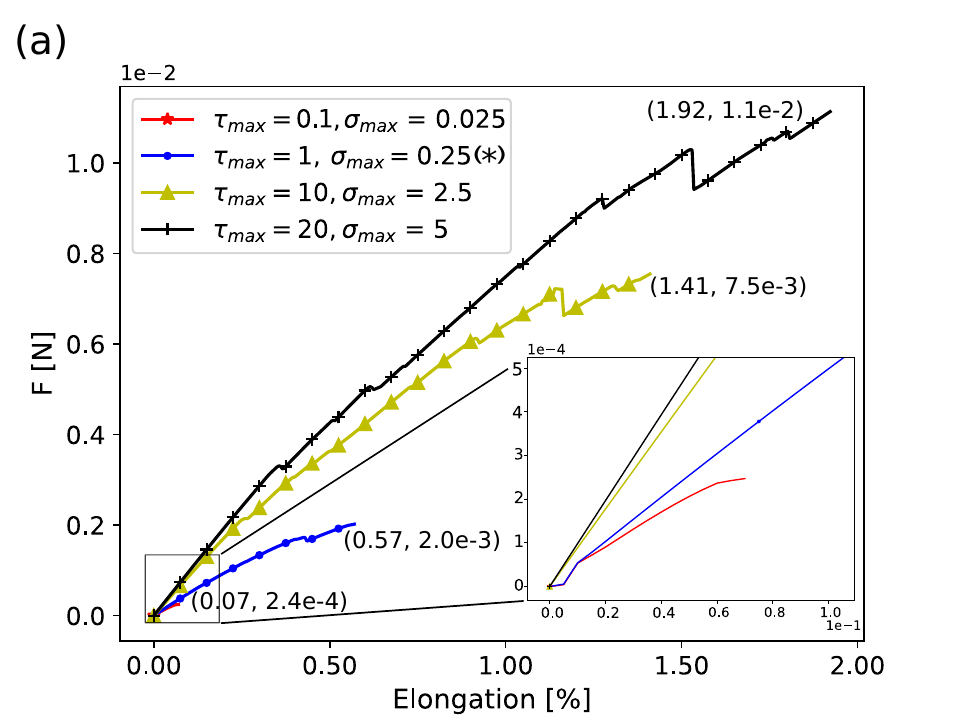}\includegraphics[width=0.5\columnwidth]{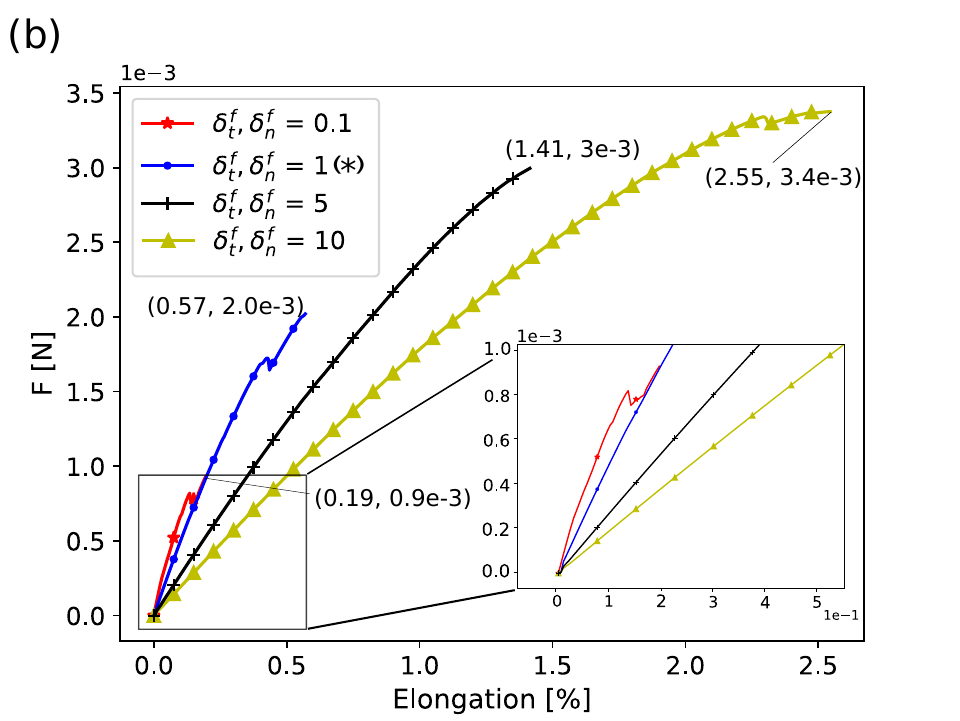}
\par\end{centering}

\caption{\label{fig:CohesiveParameterInfluence} Influence of cohesive parameters
on the force-elongation curve. (a) Varying the cohesive
strength. (b) Varying the cohesive separation. }
\end{figure}

\paragraph{Influence of cohesive model parameters}
The cohesive parameters describe the mechanical properties of the bond and the energy stored in the fiber-fiber contacts. The choice of cohesive parameters is based on fiber joint tests from available literature. The reported experimental values in a fiber-fiber shearing type study range from 0.2-16 MPa according to Magnusson et. al. \cite{magnusson2013experimental} and about 3-5 MPa for softwood and hardwood joints, according to Jajcinovic et. al. \cite{jajcinovic2016strength}, respectively. The normal cohesive strength was reported to be approximately four times smaller than that in tangential direction as given in \cite{magnusson2013numerical,borodulina2018effect,marais2014new}. Worth mentioning, the determined strength values in peeling or shearing type studies are to some extent, always mixed in the loading and due to experimental setups never in its pure form \cite{magnusson2016investigation}. Less information is available on the critical separations of fiber joints in experimental context, due to difficult experimental realizations as mentioned. As for the numerical simulation using cohesive zone models, work \cite{magnusson2016investigation} set the cohesive final separation to be 1 $\mu$m for both normal and tangential direction, and in \cite{borodulina2018effect, mansour2019stochastic} they were assumed to be 1.56 $\mu$m and 0.35 $\mu$m, respectively.
Given the variation in the reported values in the literature, a range of parameters are therefore checked to study their influence on the mechanical response of the fiber network. Based on the literature, we set the default cohesive material parameter for $\tau_{max} = 1$ MPa, $\sigma_{max} = 0.25$ MPa and follow \cite{magnusson2016investigation} setting $\delta_t^f = 1~\mu$m and $ \delta_n^f = 1 ~\mu$m, with corresponding critical separations determined by equating the energy terms $\phi_{n}=\delta_{n}^{c} \sigma_{max} \exp\left(1\right)$
and $\phi_{t}=\delta_{t}^{c} \tau_{max}\sqrt{0.5\exp\left(1\right)}$ from our work with bilinear cohesive energy $\phi_{n}= \frac{1}{2} \sigma_{max}\delta_{n}^{f}$ and $\phi_{t}=\frac{1}{2}~\tau_{max} \delta_{t}^{f}$ from \cite{magnusson2016investigation,borodulina2018effect}. The variation range is given in Tab.~\ref{tab:Parameter-study}. Generally, non-linearity in force-elongation behaviour is observed. From the simulation results, after a small elastic region, the curves become nonlinear due to the inhomogeneous damage development at different fiber bonds and the progressive and cooperative interactions between damaged interfaces. It is noted that  force drop occurs during the loading, which can be reasoned by reorientation and readjustments of the fibers under the loading. Fig.~\ref{fig:CohesiveParameterInfluence}(a) demonstrates the influence
of the parameters $\sigma_{max}, \tau_{max}$ on the mechanical behavior, where the blue curve with corresponding (*) in the legend of the figure denotes the results with default parameters. As it is shown, reducing $\tau_{max}$ by 10 times from 1 MPa to 0.1 MPa (respectively for $\sigma_{max}$), the elongation reduces by a factor of around 8 (from 0.5 8\%  to 0.07 \%), the maximal force reduces about 8.3 times as well (from 2 mN to 0.24 mN).
Increasing $\tau_{max}$ by a factor of 10 (respectively for $\sigma_{max}$), the elongation increases by a factor of around 2.4 (from 0.58 \%  to 1.41 \%), the reaction force increases around 4 times (from 2 mN to 7.5 mN). Further increase of $\tau_{max}$ from 10 MPa to 20 MPa shows an increase of 1.4 times in elongation and of 1.5 times in maximal force. Clearly, the force and elongation have non-linear responses to the change of the cohesive strength,  the factors w.r.t changing cohesive strength are rather same for both force and elongation quantities. For the effective stiffness of the sample, the secant modulus is considered (ratio of the maximal force per unit area to elongation), since not every simulated case has a clear elastic region. The effective stiffness is less sensitive against the change of the cohesive strength, which implies in turn, that the overall stiffness of the fiber network are less influenced by the cohesive strength. Fig.~\ref{fig:CohesiveParameterInfluence}(b) demonstrates the influence of the cohesive separation $\delta_n^f, \delta_t^f$ on the force-elongation curves. When reducing $\delta_{t}^{f},\delta_{n}^{f}$ from 1 $\mu$m to 0.1 $\mu$m, maximal force decreases by a factor of 2.2 (from 2 mN to 0.9 mN) and elongation decreases by a factor of 3 (from 0.57 \% to 0.19 \%). Increasing from default parameter 1 $\mu$m, to 5 $\mu$m and 10 $\mu$m, the  failure force changes from 2 mN to 3 mN and 3.4 mN, which corresponds to a increase factor of 1.5, 1.13, respectively. For the elongation, factors of 3, 2.5, 1.8 are obtained for values increasing from 0.19 \%, 0.57 \%, 1.41 \% and 2.55 \%. The change in effective stiffness (as secant modulus) shows a decreasing trend with factors of 0.74, 0.6 and 0.63. Therefore, increasing the cohesive separation, the fiber network is considered to be softer, stronger and has a higher deformability, whereas changing cohesive strength does not influence the stiffness of the fiber network, but makes it stronger and more extensible as well.

\begin{figure}[H]
\begin{centering}
\includegraphics[width=0.5\columnwidth]{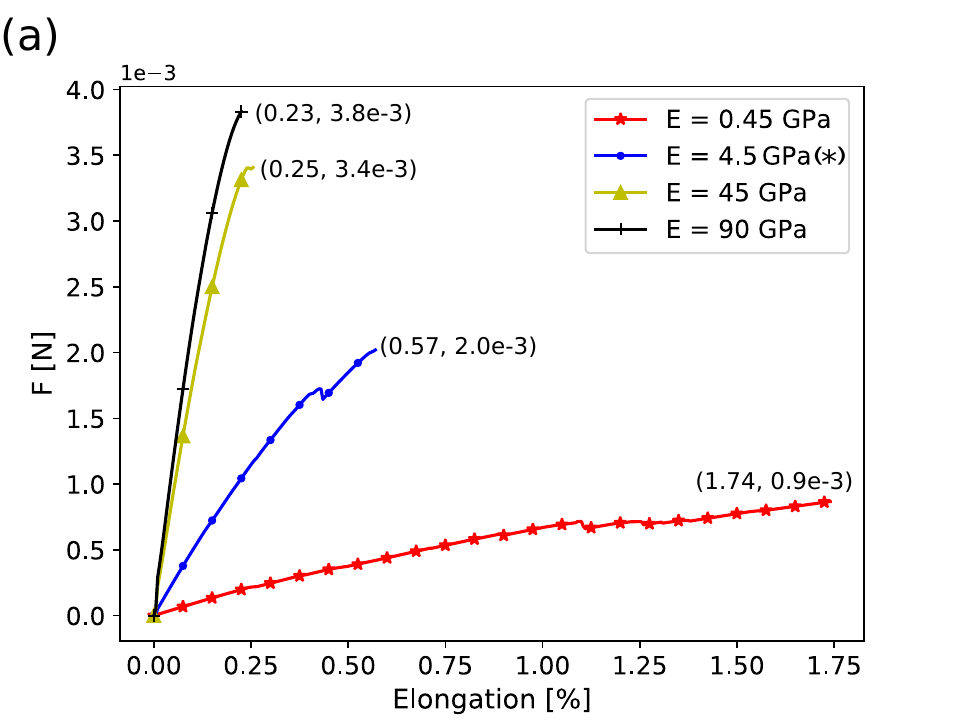}\includegraphics[width=0.5\columnwidth]{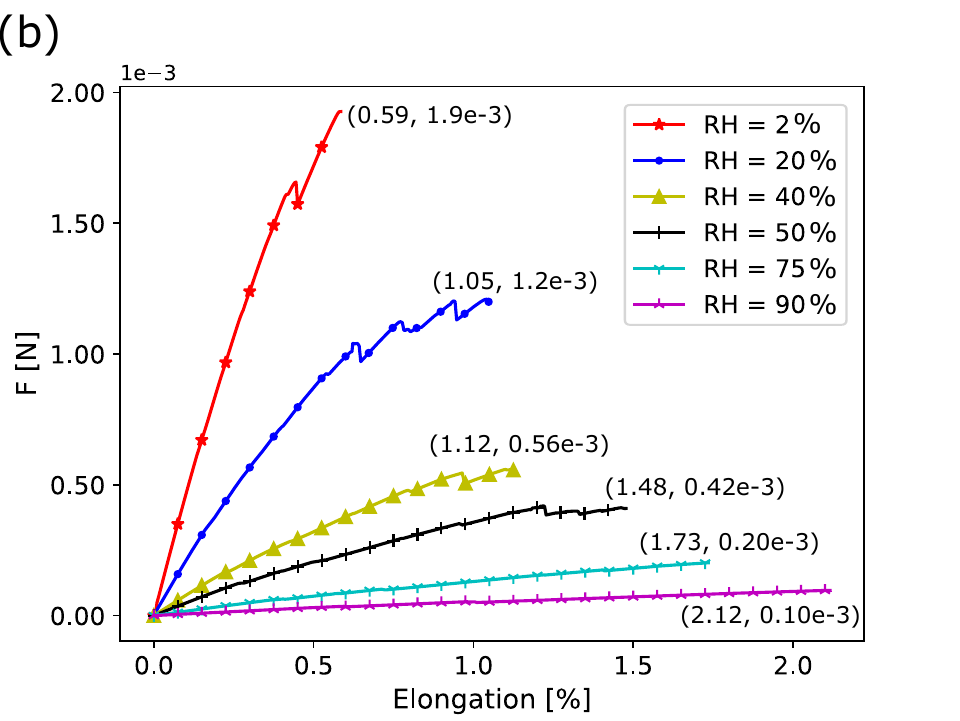}
\par\end{centering}
\caption{\label{fig:ElasticParameterInflunceAndRH} (a) Influence of the elastic modulus
on the force-elongation curve. (b) Influence of the RH on the force-elongation curve. }

\end{figure}

\paragraph{Influence of Young's modulus}
As discussed previously, our determined fiber Young's modulus appears to be lower than the reported values in the literature. We study hence here the influence of the variation of the elastic modulus on the force-elongation behaviour, given our experimentally determined results and those reported in the literature. Default elastic modulus is set to be 4.5 GPa, extrapolated from our calibrated exponential law at RH = 0 \%. Fig.~\ref{fig:ElasticParameterInflunceAndRH}(a) shows that increasing the Young's modulus by a factor of 10, the maximal force increases from 2 mN to 3.4 mN, which corresponds to a factor of 1.7.  The elongation decreases from 0.57 \% to 0.25 \% and gives a factor of 2.3. Further increasing E from 45 GPa to 90 GPa, only minor changes are observed in the mechanical responses. More interestingly, reducing 4.5 GPa to 0.45 GPa, the elongation increases by a factor of 3, and the maximal force decreases approx. 2 times. The effective stiffness changes significantly, with a factor of 1.09, 2.3 and 3 when changing 90 GPa to 45 GPa and to 4.5 GPa,  finally to 0.45 GPa. The mechanical responses scale exponentially, and implies a strong impact on the extensiablity and the effective stiffness of the fiber network by the stiffness of constituent fibers. It is also notable, that the mechanical behaviour changes from very brittle to a more ductile behaviour, with decreasing Young's modulus.

\paragraph{Influence of relative humidity}
The influence of RH on the force-elongation curves
are presented in Fig.~\ref{fig:ElasticParameterInflunceAndRH}(b).  Note that in our simulations two key parameters were subjected to the RH influence, namely the fiber Young's modulus and the cohesive strength parameters. The calibrated power law of fiber elasticity decay was used here accordingly. Further, we assumed the decaying parameter $K$ used in the humidity dependent cohesive zone model in  Eq.~\ref{CZhumidity} to be 0.3, which implies that for e.g. $\Delta$RH = 30\%, the cohesive strength is decreased by approx. 10\%. This amount of decrease is similar to the interfiber joint test subjected to varying RH observed in \cite{jajcinovic2018influence}.  
As depicted in the figure, for a reference dry state of RH = 0\%, with incremental steps varying from 2\% to 90\% , the stiffness and the maximal force of the fiber network gradually decrease, whereas the elongation or extensiability increases with RH levels.
The transition from brittle to ductile behaviour is observed here similarly. This phenomenon is in agreement with most experimental results and fundamental for paper-making process, where e.g wet pressing requires high degree of deformability and extensiablity of the material, before any breakage of the long paper during fabrication in the paper machine occurs. The mechanical behaviour subjected to RH effect is similar to the variation of elastic modulus, it shows a non-linear decreasing trend in maximal force and effective stiffness, but an improvement in elongation, and converges to the higher RH levels. As seen earlier, change of cohesive strength influences the elongation and maximal force, but to a smaller extent the effective stiffness. Also for a cohesive strength decay parameter K = 0.3 (i.e. maximal cohesive strength decrease of 30\%), minor effect would be resulted on the mechanical behaviour, in comparison with previous parameter study by decreasing a factor of 10 (1000 \%). On the other hand, the obtained E-RH behaviour is due to a exponential decay, therefore makes RH dependent elastic modulus a greater impact contributor to the RH-dependent mechanical behaviour of the underlying fiber network structure. However, since the microscopic fibrous network manifests always a certain degree of randomness, the mechanical properties are sensitive to the geometrical variations. The effect of geometrical variation and the uncertainty quantification of micro-structural features are not the focus in the current work, and have been investigated in our previous work \cite{lin2020data} using a data-driven and machine learning based approach. It must be mentioned that the mechanical behaviour of the fiber network
is also strongly size dependent as reported in work \cite{hristopulos2004structural}. The strength of the paper sample follows a weakest-link scaling law. The
density of the fiber network naturally plays an important role,  as increasing the basis weight of the paper sheet can lead to increase in its strength and stiffness as well. More details regarding these aspects can be also found in studies by \cite{kulachenko2012direct,borodulina2012stress}.

\begin{figure}[H]
\begin{centering}
\includegraphics[width=0.5\columnwidth]{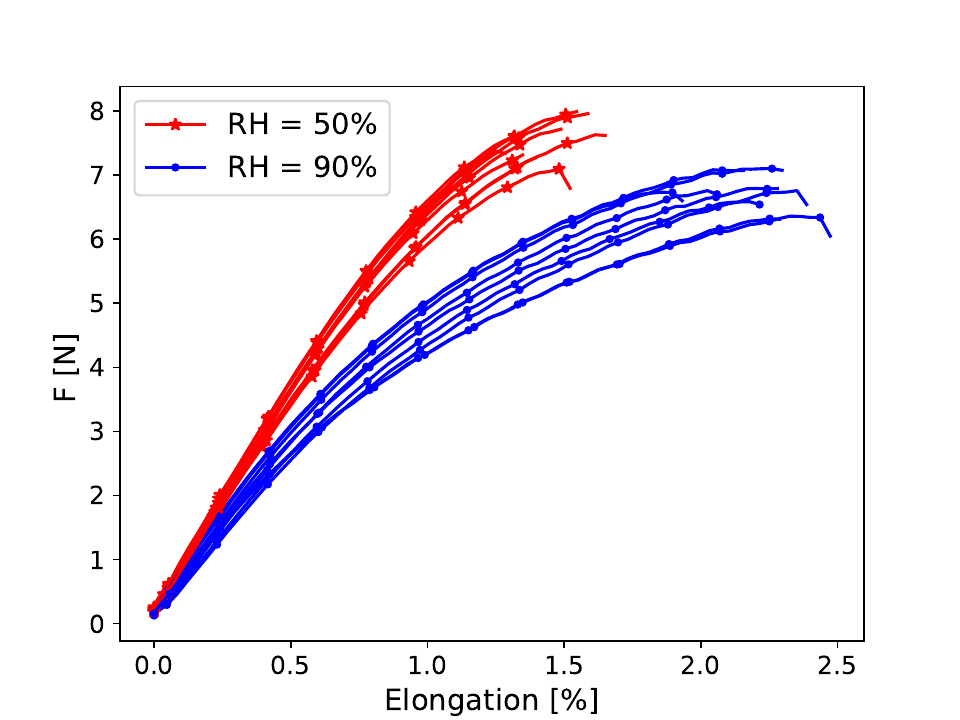}
\par\end{centering}

\caption{\label{fig:RH_paperstrip} Influence of RH
on the force-elongation curve of paper sheet samples.}
\end{figure}

\paragraph{Tensile test of paper sheet}
To provide additional information of RH influence on the mechanical behaviour, tensile test on paper sheetn samples were performed at RH = 50\% and 90\%, with the mentioned sample size in previous section. Results are shown in Fig.~\ref{fig:RH_paperstrip}. For RH = 50\%, data from nine samples were collected and the maximal force was determined to be 
$7.56\pm0.31$N, with the corresponding elongation at break to be 1.45 $\pm$ 0.12\%. For RH = 90\%, data from ten samples were collected. The maximal force was $6.6\pm 0.45$N, with the corresponding elongation $2.18\pm0.15\%$. As the figure shows Clearly, the maximal force and the effective stiffness decrease and the elongation increases with increasing RH levels. The previous simulation results agree well with the presented tensile test to demonstrate the RH influence on the mechanical behaviour in a qualitative way . Worth mentioning, it is still challenging to make a comparison in a quantitative manner, since the underlying microstructure varies locally and significantly, when larger samples are considered. From the computational perspective, to consider a larger sample in our framework, limitation in computation power arise due to number of fibers and interfiber contacts, that need to be resolved sufficiently to obtain mesh-independent results.  Therefore, further modelling techniques are required, e.g homogenisation of the presented fiber network models as an element or starting point for a multi-scale modelling paradigm. For that purpose, our bottom-up approach considering RH influence delivers important information for understanding and designing of paper materials on the fiber, and fiber network scale and serve as baseline for studying mechanical and fracture behaviour of paper material models on paper sheet, or even higher scale.

\subsection{ Failure mechanism: Fiber pull-out}
To study the failure resistance and mechanism of thin sheet fiber composites, Seth~et.~al.~\cite{seth1974fracture} investigated paper sheet by means of linear elastic fracture mechanics. Zechner~et.~al.~\cite{zechner2013determining} treated paper as a model material and carried out transverse test on double-edge notch tension specimens using digital image correlation. Recent work by Golkhosh~et.~al. \cite{golkhosh20204d} used synchrotron tomographic imaging technique and studied in-situ the failure mechanism of the fiber network in a high resolved fashion.
In the present work, using macro videography, paper sheets were locally wetted, and fibers were fluorescently stained to allow tracking of their individual movement and deformation during the successive rapture . To be in line with axis convention of the simulation condition in Sec. \ref{fiber network generation and FE simulation conditions}, loading direction is denoted as MD and the perpendicular direction is denoted as CD. However, it must be noted that the lab-engineered paper sheets are close to isotropic and therefore, the fibers do not show significant orientation alignment. MD and CD are rather axis convention in this setting.

\begin{figure}[H]
\begin{centering}
\includegraphics[width=0.9\columnwidth]{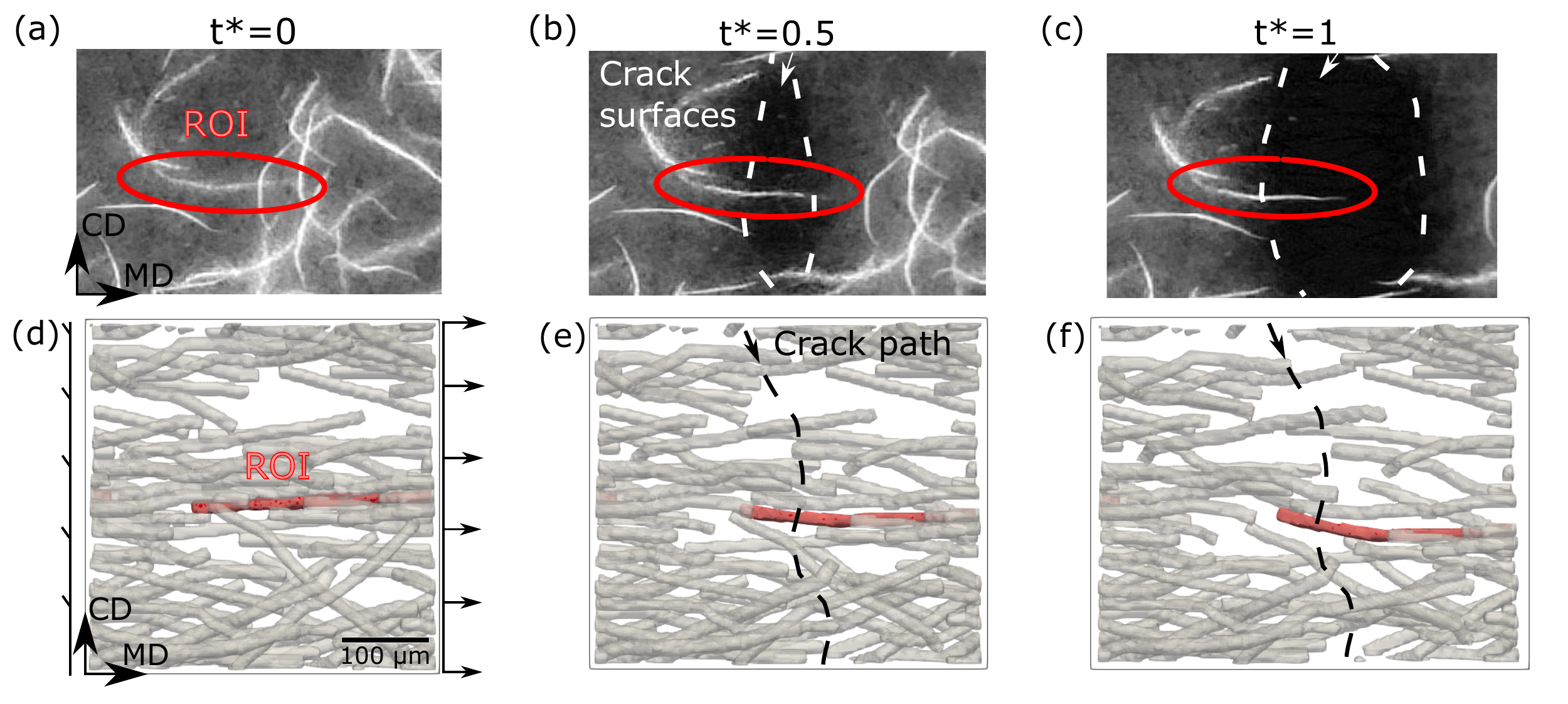}
\par\end{centering}


\caption{\label{fig:failure MD} 
Pull-out process for a fiber orientated in the MD and thus parallel to the loading direction.   
\textbf{(a-c)} Snapshots during the experimental failure test. The illuminated fiber is marked as ROI at the initial state and highlighted by the red ellipse during the tensile test. The fracture surfaces are indicated by the dashed line. (d-f): Snapshots of the deformed fiber network during the corresponding FE simulations. The fiber of interest ROI is highlighted in red, while the fracture surface is implicated by the back dashed line.}
\end{figure}

\begin{figure}[H]
\begin{centering}
\includegraphics[width=0.9\columnwidth]{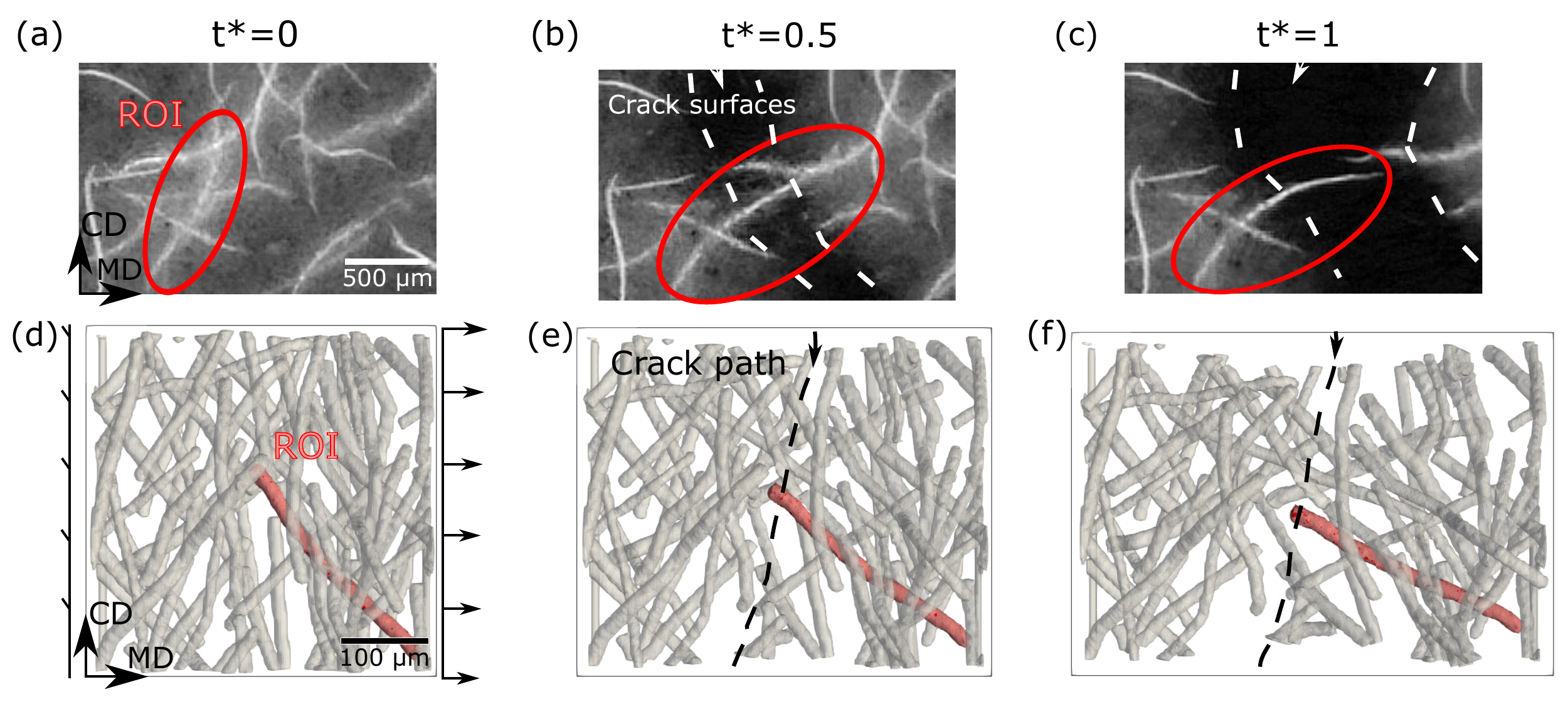}
\par\end{centering}

\caption{\label{fig: failure CD} Pull-out process for a fiber orientated in the CD and thus perpendicular to the loading direction. \textbf{(a-c)} Snapshots during the experimental failure test. The illuminated fiber is marked as ROI at the initial state and highlighted by the red ellipses during the tensile test. The fracture surfaces are indicated by the dashed line. (d-f): Snapshots of the deformed fiber network during the corresponding FE simulations. The fiber of interest ROI is highlighted in red, while the fracture surface is implicated by the back dashed line.}
\end{figure}

In the following, the considered two exemplary fiber networks with specific single fiber alignment w.r.t  MD are presented. At the same time, corresponding finite element simulations of the fiber network samples including the cohesive fiber contact model were carried out for comparison. Simulation boundary conditions as in Sec. \ref{fiber network generation and FE simulation conditions} were slightly modified, namely $u_{CD}=0$
on the front and back boundary are discarded to allow more fiber movement in the failure analysis.
The results for the tensile test with fiber alignment (The Fiber of interest is marked as region of interest (ROI)) within MD are presented in Fig.~\ref{fig:failure MD}. The experimental images in Fig.~\ref{fig:failure MD}(a) shows the initial positions of the fiber before the tensile test. As the subsequent snapshots~ Fig.~\ref{fig:failure MD}(b-c) taken during the rupture clearly demonstrate, those fibers, which lie rather parallel to the loading direction (MD in this case), are pulled out from the neighbouring fiber networks. For better understanding, the crack surfaces are indicated in the figures. No fiber breakage was observed for all the visible illuminated fibers near the crack surfaces. This is evidenced by the facts that the illuminated single fibers retain almost their initial lengths during the whole process. It implies that the fiber-fiber debonding is the main failure mechanism in this tensile test. Moreover, it is noticeable that the fibers oriented in the loading direction barely undergo any deformation or reorientation. These experimental observations are strongly supported by the fiber network simulation results shown in Fig.~\ref{fig:failure MD}(d-f). In particular, one fiber aligned in the MD is highlighted in red and tracked during the tensile test. Due to the numerical convergence issue related to the rigid body movement after progressive rupture, simulation results cannot be shown for the final fully separated stage of the fiber pull-out. But, Fig.~\ref{fig:failure MD}(e-f) show the apparent debonding of the fiber in red from the neighbouring fibers. The fiber is slightly deformed, but remains the original orientation in the MD. The scenario changes in the case of the tensile test with fiber orientated in the CD. The initial image with the illuminated fibers is presented in Fig.~\ref{fig: failure CD}(a). Again, we particularly studied the fiber marked as ROI, as the fiber is aligned perpendicular to the loading direction. Fig.~\ref{fig: failure CD}(b-c) illustrate that this fiber is progressively deformed and reoriented during the test. This can be partly explained by the bending effect resulted from the relative geometry relation between the fiber orientation and the loading direction. It can also be attributed to the bond orientation and separation process, where most of the separation are orthogonal to the loading direction. Considerable deformation and reorientation of one similar fiber has also been observed in the finite element simulations Fig.~\ref{fig: failure CD}(d-f). Both the experimental images and the simulations indicate that after the reorientation of the fiber, it is pulled out from the opposite crack surface. Our observation are in good agreement with the observation reported in~\cite{siqueira2012polyamidoamine} for the wet tensile test. However, for case of dry paper \cite{golkhosh20204d}, both fiber and fiber joint breakage are observed, whereas for paper under humid environment or in wetted regions, fiber pull-outs following interfiber joint breakage are the dominant case.

\section{Conclusion}

In summary, on different scales, starting from the fiber scale: We experimentally determined the humidity dependent longitudinal fiber elastic modulus using atomic force microscopy and proposed an power law for decaying of the humidity dependent fiber elasticity. The obtained Young's modulus was about 4 GPa and slightly lower than values published in the literature. It could be attributed to the fact that the tested fiber was rather a fiber from a finished paper sheets than a raw one without the influence of a paper-making process.  Hygroscopic expansion coefficient was obtained to be 0.35 using confocal laser scanning microscopy, by recording the change of fiber cross-section swelling at different relative humidity levels.  On the fiber/fiber cross scale: humidity dependent cohesive zone model was proposed and applied,  jointly with the humidity dependent elasticity to simulate the mechanical behaviour of fiber network under tensile test on a higher scale. On the fiber network scale: we carried out a series of parameter studies, and determined that key contribution to RH influence on the mechanical behaviour is due to the RH-dependent exponential decaying from single fiber elasticity, whereas the interfiber properties play a minor role in the reported material parameter range. The mechanical behaviour of simulated tensile test under humidity influence was in good qualitative agreement, compared to our experimental test on the paper sheet scale. The fiber network failure in a locally wetted region was revealed by tracking individual fluorescently stained fibers using fluorescence macro videography. Compared to the simulated fiber network failure, both experimental tensile test and the cohesive finite element simulations demonstrated the pull-out of fibers and implied the significant role of the fiber-fiber debonding in the failure process of the wet paper. 

\newpage
\section{CRediT authorship statement}

\textbf{B. Lin:} Conceptualization, Methodology, Data Curation, Software, Formal analysis, Visualization, Writing - Original draft. 
\textbf{J. Auernhammer:}  
 Data Curation, Visualization, Writing - Review \& Editing. \textbf{J.  Sch\"afer:} Data Curation, Visualization, Writing - Review \& Editing.   \textbf{T. Meckel:} Data Curation, Visualization. \textbf{B. Xu:} Conceptualization, Formal analysis, Writing - Review \& Editing, Supervision, Funding acquisition.  \textbf{R. Stark:} Supervision, Funding acquisition.
\textbf{M. Biesalski:}  Supervision, Funding acquisition.

\section{Declaration of competing interest} 
The authors declare that they have no known competing financial interests or personal relationships that could have appeared to influence the work reported in this paper.

\section{Acknowledgements}

This work was supported by German Research Foundation (DFG). B. Lin acknowledges the financial support under
the grant agreement No. 405422877 of the
Paper Research project (FiPRe) and thanks Y. Yang for his help on HPC settings. J. Auernhammer acknowledges DFG for the financial funding under No. 405549611 and thanks Lars-Oliver Heim for the modification of the cantilever with the colloidal probe. J. Sch\"afer acknowledges DFG for the financial funding under No. 405422473.  The authors also greatly appreciate
their access to the Lichtenberg High Performance Computer and the
technical supports from the HHLR, Technical university Darmstadt.





\newpage
\section*{References}

\addcontentsline{toc}{section}{\refname}\bibliography{reference}

\begin{thebibliography}{10}

\bibitem{gong2017turning}
Max~M Gong and David Sinton.
\newblock Turning the page: advancing paper-based microfluidics for broad
  diagnostic application.
\newblock {\em Chemical reviews}, 117(12):8447--8480, 2017.

\bibitem{SHEN2019389}
Liu-Liu Shen, Gui-Rong Zhang, Tizian Venter, Markus Biesalski, and Bastian~J.M.
  Etzold.
\newblock Towards best practices for improving paper-based microfluidic fuel
  cells.
\newblock {\em Electrochimica Acta}, 298:389 -- 399, 2019.

\bibitem{schabel2019role}
Samuel Schabel and Markus Biesalski.
\newblock The role of paper chemistry and paper manufacture in the design of
  paper-based diagnostics.
\newblock In {\em Paper-based Diagnostics}, pages 23--46. Springer, 2019.

\bibitem{liu2014filter}
Mingkai Liu, Sixin He, Wei Fan, Yue-E Miao, and Tianxi Liu.
\newblock Filter paper-derived carbon fiber/polyaniline composite paper for
  high energy storage applications.
\newblock {\em Composites science and technology}, 101:152--158, 2014.

\bibitem{pantaloni2021interfacial}
Delphin Pantaloni, Anton~Lo{\"\i}c Rudolph, Darshil~U Shah, Christophe Baley,
  and Alain Bourmaud.
\newblock Interfacial and mechanical characterisation of biodegradable
  polymer-flax fibre composites.
\newblock {\em Composites Science and Technology}, 201:108529, 2021.

\bibitem{regazzi2019microstructural}
Arnaud Regazzi, Maxime Teil, Pierre~JJ Dumont, Barthelemy Harthong, Didier
  Imbault, Robert Peyroux, and Jean-Luc Putaux.
\newblock Microstructural and mechanical properties of biocomposites made of
  native starch granules and wood fibers.
\newblock {\em Composites Science and Technology}, 182:107755, 2019.

\bibitem{lee2006biodegradable}
Seung-Hwan Lee and Siqun Wang.
\newblock Biodegradable polymers/bamboo fiber biocomposite with bio-based
  coupling agent.
\newblock {\em Composites Part A: applied science and manufacturing},
  37(1):80--91, 2006.

\bibitem{salmen1980moisture}
N~Lennart Salmen and Ernst~L Back.
\newblock Moisture-dependent thermal softening of paper, evaluated by its
  elastic modulus.
\newblock {\em Tappi}, 63(6):117--120, 1980.

\bibitem{jajcinovic2018influence}
Marina Jajcinovic, Wolfgang~J Fischer, Andreas Mautner, Wolfgang Bauer, and
  Ulrich Hirn.
\newblock Influence of relative humidity on the strength of hardwood and
  softwood pulp fibres and fibre to fibre joints.
\newblock {\em Cellulose}, 25(4):2681--2690, 2018.

\bibitem{tejado2010does}
Alvaro Tejado and Theo~GM van~de Ven.
\newblock Why does paper get stronger as it dries?
\newblock {\em Materials today}, 13(9):42--49, 2010.

\bibitem{neimo1999papermaking}
L.~Neimo, Suomen~Paperi insin{\"o}{\"o}rien Yhdistys, Technical~Association
  of~the Pulp, and Paper Industry.
\newblock {\em Papermaking Chemistry}.
\newblock Number Buch 4 in Papermaking chemistry. Fapet Oy, 1999.

\bibitem{britt1948review}
KW~Britt.
\newblock Review of developments in wet strength paper.
\newblock In {\em Technical association papers}, volume~31, pages 594--596,
  1948.

\bibitem{hubbe2008cellulosic}
Martin~A Hubbe, Orlando~J Rojas, Lucian~A Lucia, and Mohini Sain.
\newblock Cellulosic nanocomposites: a review.
\newblock {\em BioResources}, 3(3):929--980, 2008.

\bibitem{gamstedt2016moisture}
E~Kristofer Gamstedt.
\newblock Moisture induced softening and swelling of natural cellulose fibres
  in composite applications.
\newblock In {\em 37th Riso International Symposium on Materials
  Science-Understanding Performance of Composite Materials-Mechanisms
  Controlling Properties, SEP 05-08, 2016, Riso, DENMARK}, 2016.

\bibitem{joffre2013swelling}
Thomas Joffre, Erik~LG Wernersson, Arttu Miettinen, Cris L~Luengo Hendriks, and
  E~Kristofer Gamstedt.
\newblock Swelling of cellulose fibres in composite materials: constraint
  effects of the surrounding matrix.
\newblock {\em Composites science and technology}, 74:52--59, 2013.

\bibitem{neagu2005influence}
R~Cristian Neagu, E~Kristofer Gamstedt, and Mikael Lindstr{\"o}m.
\newblock Influence of wood-fibre hygroexpansion on the dimensional instability
  of fibre mats and composites.
\newblock {\em Composites Part A: Applied Science and Manufacturing},
  36(6):772--788, 2005.

\bibitem{motamedian2019simulating}
Hamid~Reza Motamedian and Artem Kulachenko.
\newblock Simulating the hygroexpansion of paper using a 3d beam network model
  and concurrent multiscale approach.
\newblock {\em International Journal of Solids and Structures}, 161:23--41,
  2019.

\bibitem{simon2020review}
Jaan-Willem Simon.
\newblock A review of recent trends and challenges in computational modeling of
  paper and paperboard at different scales.
\newblock {\em Archives of Computational Methods in Engineering}, pages 1--20,
  2020.

\bibitem{borodulina2018effect}
Svetlana Borodulina, Hamid~Reza Motamedian, and Artem Kulachenko.
\newblock Effect of fiber and bond strength variations on the tensile stiffness
  and strength of fiber networks.
\newblock {\em International Journal of Solids and Structures}, 154:19--32,
  2018.

\bibitem{Li:730091}
Yujun Li, Zengzhi Yu, Stefanie Reese, and Jaan-Willem Simon.
\newblock {E}valuation of the out-of-plane response of fiber networks with a
  representative volume element model.
\newblock {\em Tappi journal}, 17(6):329--339, 2018.

\bibitem{gross2009technische}
D.~Gross, W.~Hauger, J.~Schrader, and W.A. Wall.
\newblock {\em Technische Mechanik}.
\newblock Number Bd. 2 in Springer-Lehrbuch. Springer Berlin Heidelberg, 2009.

\bibitem{joffre2016method}
Thomas Joffre, Per Isaksson, Pierre~JJ Dumont, S~Rolland Du~Roscoat, Simon
  Sticko, Laurent Org{\'e}as, and E~Kristofer Gamstedt.
\newblock A method to measure moisture induced swelling properties of a single
  wood cell.
\newblock {\em Experimental mechanics}, 56(5):723--733, 2016.

\bibitem{schindelin2012fiji}
Johannes Schindelin, Ignacio Arganda-Carreras, Erwin Frise, Verena Kaynig, Mark
  Longair, Tobias Pietzsch, Stephan Preibisch, Curtis Rueden, Stephan Saalfeld,
  Benjamin Schmid, et~al.
\newblock Fiji: an open-source platform for biological-image analysis.
\newblock {\em Nature methods}, 9(7):676--682, 2012.

\bibitem{magnusson2013numerical}
Mikael~S Magnusson and S{\"o}ren {\"O}stlund.
\newblock Numerical evaluation of interfibre joint strength measurements in
  terms of three-dimensional resultant forces and moments.
\newblock {\em Cellulose}, 20(4):1691--1710, 2013.

\bibitem{mcgarry2014potential}
J~Patrick McGarry, {\'E}amonn~{\'O} M{\'a}irt{\'\i}n, Guillaume Parry, and
  Glenn~E Beltz.
\newblock Potential-based and non-potential-based cohesive zone formulations
  under mixed-mode separation and over-closure. part i: Theoretical analysis.
\newblock {\em Journal of the Mechanics and Physics of Solids}, 63:336--362,
  2014.

\bibitem{jemblie2017review}
Lise Jemblie, Vigdis Olden, and Odd~Magne Akselsen.
\newblock A review of cohesive zone modelling as an approach for numerically
  assessing hydrogen embrittlement of steel structures.
\newblock {\em Philosophical Transactions of the Royal Society A: Mathematical,
  Physical and Engineering Sciences}, 375(2098):20160411, 2017.

\bibitem{park2012computational}
Kyoungsoo Park and Glaucio~H Paulino.
\newblock Computational implementation of the ppr potential-based cohesive
  model in abaqus: Educational perspective.
\newblock {\em Engineering fracture mechanics}, 93:239--262, 2012.

\bibitem{permann2020moose}
Cody~J. Permann, Derek~R. Gaston, David Andr{\v{s}}, Robert~W. Carlsen, Fande
  Kong, Alexander~D. Lindsay, Jason~M. Miller, John~W. Peterson, Andrew~E.
  Slaughter, Roy~H. Stogner, and Richard~C. Martineau.
\newblock {MOOSE}: Enabling massively parallel multiphysics simulation.
\newblock {\em {SoftwareX}}, 11:100430, 2020.

\bibitem{kulachenko2012direct}
Artem Kulachenko and Tetsu Uesaka.
\newblock Direct simulations of fiber network deformation and failure.
\newblock {\em Mechanics of Materials}, 51:1--14, 2012.

\bibitem{geuzaine2009gmsh}
Christophe Geuzaine and Jean-Fran{\c{c}}ois Remacle.
\newblock Gmsh: A 3-d finite element mesh generator with built-in pre-and
  post-processing facilities.
\newblock {\em International journal for numerical methods in engineering},
  79(11):1309--1331, 2009.

\bibitem{lin2020data}
Binbin Lin, Yang Bai, and Bai-Xiang Xu.
\newblock Data-driven microstructure sensitivity study of fibrous paper
  materials.
\newblock {\em Materials \& Design}, page 109193, 2020.

\bibitem{blechschmidt2013taschenbuch}
J{\"u}rgen Blechschmidt.
\newblock {\em Taschenbuch der Papiertechnik}.
\newblock Carl Hanser Verlag GmbH Co KG, 2013.

\bibitem{janko2015cross}
Marek Janko, Michael Jocher, Alexander Boehm, Laura Babel, Steven Bump, Markus
  Biesalski, Tobias Meckel, and Robert~W Stark.
\newblock Cross-linking cellulosic fibers with photoreactive polymers:
  visualization with confocal raman and fluorescence microscopy.
\newblock {\em Biomacromolecules}, 16(7):2179--2187, 2015.

\bibitem{quesada2011nanomechanical}
Raul Quesada~Cabrera, Filip Meersman, Paul~F McMillan, and Vladimir Dmitriev.
\newblock Nanomechanical and structural properties of native cellulose under
  compressive stress.
\newblock {\em Biomacromolecules}, 12(6):2178--2183, 2011.

\bibitem{fernando2017mechanical}
S~Fernando, CF~Mallinson, C~Phanopolous, DA~Jesson, and JF~Watts.
\newblock Mechanical characterisation of fibres for engineered wood products: a
  scanning force microscopy study.
\newblock {\em Journal of Materials Science}, 52(9):5072--5082, 2017.

\bibitem{eichhorn2001young}
SJ~Eichhorn and RJ~Young.
\newblock The young's modulus of a microcrystalline cellulose.
\newblock {\em Cellulose}, 8(3):197--207, 2001.

\bibitem{kompella2002micromechanical}
Mohan~K Kompella and John Lambros.
\newblock Micromechanical characterization of cellulose fibers.
\newblock {\em Polymer Testing}, 21(5):523--530, 2002.

\bibitem{miyake2000tensile}
Hajime Miyake, Yasuo Gotoh, Yutaka Ohkoshi, and Masanobu Nagura.
\newblock Tensile properties of wet cellulose.
\newblock {\em Polymer journal}, 32(1):29--32, 2000.

\bibitem{neagu2007modelling}
R~Cristian Neagu and E~Kristofer Gamstedt.
\newblock Modelling of effects of ultrastructural morphology on the
  hygroelastic properties of wood fibres.
\newblock {\em Journal of Materials Science}, 42(24):10254--10274, 2007.

\bibitem{almgren2009moisture}
Karin~M Almgren, E~Kristofer Gamstedt, Fredrik Berthold, and Mikael
  Lindstr{\"o}m.
\newblock Moisture uptake and hygroexpansion of wood fiber composite materials
  with polylactide and polypropylene matrix materials.
\newblock {\em Polymer composites}, 30(12):1809--1816, 2009.

\bibitem{magnusson2013experimental}
Mikael~S Magnusson, Xiaobo Zhang, and S{\"o}ren {\"O}stlund.
\newblock Experimental evaluation of the interfibre joint strength of
  papermaking fibres in terms of manufacturing parameters and in two different
  loading directions.
\newblock {\em Experimental mechanics}, 53(9):1621--1634, 2013.

\bibitem{jajcinovic2016strength}
Marina Jajcinovic, Wolfgang~J Fischer, Ulrich Hirn, and Wolfgang Bauer.
\newblock Strength of individual hardwood fibres and fibre to fibre joints.
\newblock {\em Cellulose}, 23(3):2049--2060, 2016.

\bibitem{marais2014new}
Andrew Marais, Mikael~S Magnusson, Thomas Joffre, Erik~LG Wernersson, and Lars
  W{\aa}gberg.
\newblock New insights into the mechanisms behind the strengthening of
  lignocellulosic fibrous networks with polyamines.
\newblock {\em Cellulose}, 21(6):3941--3950, 2014.

\bibitem{magnusson2016investigation}
Mikael~S Magnusson.
\newblock Investigation of interfibre joint failure and how to tailor their
  properties for paper strength.
\newblock {\em Nordic Pulp \& Paper Research Journal}, 31(1):109--122, 2016.

\bibitem{mansour2019stochastic}
Rami Mansour, Artem Kulachenko, Wei Chen, and M{\aa}rten Olsson.
\newblock Stochastic constitutive model of isotropic thin fiber networks based
  on stochastic volume elements.
\newblock {\em Materials}, 12(3):538, 2019.

\bibitem{hristopulos2004structural}
Dionissios~T Hristopulos and Tetsu Uesaka.
\newblock Structural disorder effects on the tensile strength distribution of
  heterogeneous brittle materials with emphasis on fiber networks.
\newblock {\em Physical Review B}, 70(6):064108, 2004.

\bibitem{borodulina2012stress}
Svetlana Borodulina, Artem Kulachenko, Mikael Nyg{\aa}rds, and Sylvain Galland.
\newblock Stress-strain curve of paper revisited.
\newblock {\em Nordic pulp \& paper research journal}, 27(2):318--328, 2012.

\bibitem{seth1974fracture}
RS~Seth and DH~Page.
\newblock Fracture resistance of paper.
\newblock {\em Journal of Materials Science}, 9(11):1745--1753, 1974.

\bibitem{zechner2013determining}
J~Zechner, M~Janko, and O~Kolednik.
\newblock Determining the fracture resistance of thin sheet fiber
  composites--paper as a model material.
\newblock {\em Composites science and technology}, 74:43--51, 2013.

\bibitem{golkhosh20204d}
F~Golkhosh, Y~Sharma, DM~Martinez, PD~Lee, W~Tsai, L~Courtois, DS~Eastwood, and
  AB~Phillion.
\newblock 4d synchrotron tomographic imaging of network and fibre level
  micromechanics in softwood paper.
\newblock {\em Materialia}, 11:100680, 2020.

\bibitem{siqueira2012polyamidoamine}
Eder~Jos{\'e} Siqueira.
\newblock {\em Polyamidoamine epichlorohydrin-based papers: mechanisms of wet
  strength development and paper repulping}.
\newblock PhD thesis, Grenoble, 2012.

\end{thebibliography}

\section*{--------------------}
\end{document}